\documentclass[authoryear]{elsarticle}
\pdfoutput=1
\usepackage{natbib}
\usepackage{amsmath}
\def\aap{A\&A}%
\def\jcap{JCAP}
\def\apj{ApJ}%
\def\apjl{ApJ}%
\def\ssr{Space~Sci.~Rev.}%

\def\half{\ensuremath{1/2}}

\newcommand{\nucl}[3]{
\ensuremath{
\phantom{\ensuremath{^{#1}_{#2}}}
\llap{\ensuremath{^{#1}}}
\llap{\ensuremath{_{\rule{0pt}{.75em}#2}}}
\mbox{#3}
}
}

\begin{document}


\begin{frontmatter}

\title{PICARD: A novel code for the Galactic Cosmic Ray propagation problem}
\author[ibk_ap]{R.~Kissmann\corref{cor_auth}}
\ead{ralf.kissmann@uibk.ac.at}

\address[ibk_ap]{Institut f\"ur Astro- und Teilchenphysik,
  Leopold-Franzens-Universit\"at Innsbruck, A-6020 Innsbruck, Austria}

\cortext[cor_auth]{Corresponding author}

\begin{abstract} In this manuscript we present a new approach for the
  numerical solution of the Galactic Cosmic Ray propagation
  problem. We introduce a method using advanced contemporary numerical
  algorithms while retaining the general complexity of other
  established codes. In this paper we present the underlying numerical
  scheme in conjunction with tests showing the correctness of the
  scheme. Finally we show the solution of a first example propagation
  problem using the new code to show its applicability to Galactic
  Cosmic Ray propagation.
\end{abstract}

\end{frontmatter}

\section{Introduction}
The Galactic Cosmic Ray propagation problem, i.e., the question how
Cosmic Rays are transported from their sources to arbitrary locations
in the Galaxy, becomes ever more relevant with recent advances in
observational techniques. Such observations yield the flux of primary
Cosmic Rays \citep[see, e.g.,][]{Mewaldt2001SSRv99_27M,
  SanukiEtAl2000ApJ545_1135S, AdrianiEtAl2011Sci332_69A,
  AdrianiEtAl2011PhysRevLett106_201101} or also of secondaries at
Earth. For neutral secondary particles also directional information
can be extracted from the data \citep[see,
  e.g.,][]{AckermannEtAl2012ApJ750_3A}. Together with a physical
description of the transport process of Cosmic Rays these data should
allow a better understanding of the physics involved in Cosmic Ray
transport.

The transport of Galactic Cosmic Rays is a diffusion-loss problem
\citep[see][]{StrongEtAl2007ARNPS57_285}. That is we have to find a
solution of the partial differential equation:
\begin{align}
  \label{EqPropagation}
  \frac{\partial \psi}{\partial t}
  -
  \nabla \cdot\left( \mathcal{D} \nabla \psi\right)
  &+ 
  \nabla \cdot \left(\vec{u} \psi\right)
  -
  \frac{\partial}{\partial p}
  \left(
  p^2 D_{pp}\frac{\partial}{\partial p}\frac{\psi}{p^2}
  \right)
  \\
  \nonumber
  &+
  \frac{\partial}{\partial p}
  \left(
  \dot p \psi - \frac{p}{3}\left(\nabla\cdot\vec{u}\right) \psi
  \right)
  =
  s(\vec{r}, p, t)
  - 
  \frac{1}{\tau} \psi
\end{align}
with
\begin{equation}
  \frac{1}{\tau} = \frac{1}{\tau_f} + \frac{1}{\tau_r}
\end{equation}
where the first term on the right hand side represents the sources of
Cosmic Ray species $\psi$, the second term gives the spatial
diffusion, the third represents the energy losses and the fourth term
gives losses by fragmentation and radioactive decay for the current
Cosmic Ray species.

This partial differential equation has been solved using different
numerical codes or analytical approximations or a mixture of both. Use
of analytical solutions or approximations within a numerical code
decreases the numerical cost to find a solution and gives more direct
idea of the underlying dependence of the solution on different
parameters. Analytical methods, however, are not suited to investigate
the Cosmic Ray propagation problem in a realistic environment, i.e.,
an environment, where all functions that determine the final outcome
of Eq. (\ref{EqPropagation}) are allowed to vary arbitrarily in
configuration- and momentum-space.

With the increasing precision of Galactic Cosmic Ray measurements an
analytical approach is far from being able to explain the fine details
in the measurements. Also a discussion of $>1$ TeV Cosmic Rays would
necessitate consideration of the Cosmic Ray transport from individual
sources. Therefore we will only discuss fully numerical methods in the
paper, thus also omitting references to such numerical codes like
\textsc{Usine} \citep[see][]{PutzeEtAl2010AnA516_A66}) that use
analytical approximations to improve the performance of the code. Such
codes aim at finding the best values for the variables in
Eq. (\ref{EqPropagation}) which, however, are assumed constant in the
space.

For the full numerical solution of the Galactic Cosmic Ray propagation
problem there are mainly two publicly available codes:
\textsc{Galprop} \citep[see][]{StrongEtAl2011Manual} and
\textsc{Dragon} \citep[see][]{EvoliEtAl2008JCAP10_18}.
\textsc{Galprop} is a very sophisticated framework that tries to
include all relevant physics for the propagation problem with a high
complexity. The \textsc{Dragon} code emerged from an earlier
\textsc{Galprop} version and has been continuously enhanced since. In
particular \textsc{Dragon} allows a significantly more complex
description for some of the transport parameters, like e.g. fully
anisotropic spatial diffusion, than currently available in
\textsc{Galprop} -- see, e.g., \citet{GaggeroEtAl2013arXiv1306_6850},
where also the effort in establishing the transition to spatially
three-dimensional simulations is shown. There are indeed some issues
with the representation of the physical parameters in \textsc{Galprop}
as is discussed in \citet{KissmannEtAl2012AIPC1505_450K}. This will
not be subject of the present paper. Here we rather diagnose the
problem that there was far less attention directed to the numerical
solver in \textsc{Galprop} than to other aspects of the code. This led
to the fact that the solver is rather outdated regarding the numerical
methods employed.

Consequently we will discuss the implementation of an up to date
numerical solver within a code that can adopt the same transport
parameters as \textsc{Galprop}, using initialisation via
\textsc{Galdef} files. In Sect.  \ref{SecNumerics} we describe the new
numerical scheme. Corresponding tests will be discussed in
Sect. \ref{SecTests} and we will show a typical example of a Galactic
Cosmic Ray propagation problem in Sec. \ref{SecExample}. Finally we
will conclude with an outlook on ongoing development of the code.

\section{A new numerical approach}
\label{SecNumerics}

As mentioned in the introduction the presently most widely used code
for the solution of the Galactic Cosmic Ray transport problem is
\textsc{Galprop}. This code was introduced some 20 years ago
\citep[see][]{StrongYoussefi1995ICRC3_48S} where the numerical solver
has only been marginal altered since that time.

The solution in the \textsc{Galprop} code is computed from a
Crank-Nicolson discretisation of the partial differential equation
Eq. \eqref{EqPropagation}, where the authors use operator splitting by
which they can apply the updating scheme to each spatial or momentum
dimension separately. To avoid the problem of having to solve a
prohibitively large amount of timesteps \textsc{Galprop} additionally
uses a procedure where a range of different timestep sizes is used for
the time integration beginning with very large steps and ending at a
user-specified smallest timestep. By this the solution can reach a
steady state faster than for a constant timestep method \citep[for
  further details see the appendix
  of][]{StrongMoskalenko1998APJ509_212}.

This solution scheme, however, has some severe shortcomings. The first
issue is that the numerical integration scheme depends on parameters
to be set by the user. Such parameters are, e.g., the largest and
smallest timestep, and the number each timestep size is supposed to be
used for the integration. The final solution of a simulation then
depends on the correct choice of these parameters. While the standard
parameters might suffice for the standard \textsc{Galprop} runs a
significant change in the configuration might lead to the necessity to
come up with a corresponding new set of integration parameters.  To
investigate the steady state solution that has been found
\textsc{Galprop} offers some diagnostic tools. These, however, have
explicitly to be administered and also interpreted by the user.
Therefore, when finding new time-integration parameters, several
simulations will have to be done with different parameters until it is
certain that a steady state is reached from a comparison of the
results.


Most of these issues arise from the fact that a time integration
scheme is used where a steady state solution is searched
for. Therefore, we are using two different approaches depending on the
question whether the parameters in Eq. \eqref{EqPropagation} are time
dependant or not. In the former case the solution is obtained by
integrating Eq. (\ref{EqPropagation}) from some initial conditions up
to the time of interest. Whenever the source term $s(\vec{r},p,t)$,
the diffusion tensor $\mathcal{D}$, the momentum loss rate $\dot p$
and the catastrophic loss times $\tau$ are time independent we are
using a solver that yields a steady state solution without any
integration in time instead. In this section we will discuss both
approaches - keeping in mind that it is also a viable option to use a
steady state solution to compute an initial condition for the time
dependant problem.

Looking at the transport equation Eq. \eqref{EqPropagation} shows that
when re-acceleration is not taken into account we only have to deal
with first order derivatives in momentum space. If additionally the
energy changes universally occur in the same direction, the momentum
space transport problem becomes particularly simple. This motivates
the choice of a dedicated solver. Even though this might seem to be a
special case it is a very common application in Galactic Cosmic Ray
propagation simulations. By comparison with the more general solver we
will later find that the solver adapted to this particular case is
indeed more efficient than the general one. In the following we will
refer to the different solvers as the \emph{re-acceleration scheme}
for the general solver and the \emph{energy-loss scheme} for the
special case without re-acceleration. We will introduce adapted
solvers for both situations.  We start by discussing the steady state
problem.

\subsection{The steady state problem}
\label{SubSecNumSteady}
Looking for a steady state solution of Eq. (\ref{EqPropagation}) means
that we are looking for a solution where the time-derivative goes to
zero. Therefore a popular option is to do a time integration instead
and integrate until a steady state solution is reached, i.e. until the
solution does not change anymore. Depending on the choice of variables
in Eq. (\ref{EqPropagation}) this, however, can take quite some time
and also a good criterion is needed to check whether the code has
found a steady state solution. In particular, one has to ask how small
a change would have to be in order to indicate such a steady
state. This is a particular issue for Cosmic Ray transport were the
solution varies over orders in magnitude especially as a function of
energy.

Here we are therefore using a different approach, where we explicitly
make use of the fact that the time derivative is supposed to be
zero. That is we are solving the equation
\begin{align}
  \label{EqPropagationSteady}
  -
  \nabla \cdot\left( \mathcal{D} \nabla \psi\right)
  &+ 
  \nabla \cdot \left(\vec{u} \psi\right)
  -
  \frac{\partial}{\partial p}
  \left(
  p^2 D_{pp}\frac{\partial}{\partial p}\frac{\psi}{p^2}
  \right)
  \\
  \nonumber
  &+
  \frac{\partial}{\partial p}
  \left(
  \dot p \psi - \frac{p}{3}\left(\nabla\cdot\vec{u}\right) \psi
  \right)
  + 
  \frac{1}{\tau} \psi
  =
  s(\vec{r}, p, t)
\end{align}

instead. For this equation it is not possible to use
dimensional splitting anymore like, e.g., employed in the \textsc{Galprop}
code. Solving Eq. (\ref{EqPropagationSteady}) requires solving the
whole equation at once. To find a numerical solution we need to
discretise this equation on a grid -- here we use the same approach as
in \textsc{Galprop}, i.e., a linear spatial grid and a logarithmic
grid in momentum space. Using such a discretisation the above PDE is
transformed into a coupled system of algebraic equations. In 1D such a
system can directly be solved by inverting the corresponding matrix
(that motivates the dimensional splitting used, e.g., in
\textsc{Galprop}), which usually is just a tridiagonal matrix. In the
present case with three spatial and one momentum dimension, however, a
direct solution is not efficient to compute anymore.

Therefore we are using an iterative method that relies heavily on the
application of multigrid methods, which turned out to lead to
excellent convergence in this case. As indicated above, we will now
discuss two different implementations of the numerical solver.

\subsection{Energy-loss Scheme}
Neglecting re-acceleration and provided that energy losses always
dominate gains by adiabatic energy changes it is possible to derive an
extremely efficient solution scheme for the Cosmic Ray transport
problem. Due to the fact that spatial advection can only be treated
using this scheme, when adiabatic energy gains are sufficiently small
everywhere, we will allow spatial advection exclusively in the full
scheme that also allows for re-acceleration. In the steady state
energy-loss scheme, we solve the reduced transport equation:
\begin{equation}
  \label{EqPropagationSteadyEnLoss}
  -\nabla \cdot \mathcal{D} \nabla \psi
  +
  \frac{\partial \dot p \psi}{\partial p} 
  + 
  \frac{1}{\tau} \psi
  =
  s(\vec{r}, p, t)
\end{equation}
We now discuss the ingredients for the particular solver
individually. First we start by discussing the corresponding
discretisation.

\subsubsection{Discretisation}
\label{SecDiscMomLoss}
With regard to the discretisaion of the steady state form of the
energy-loss transport equation given in
Eq. (\ref{EqPropagationSteadyEnLoss}) we use different approaches for
momentum and configuration space.

Instead of using a finite difference discretisation we use an
integration in momentum space to discretise the momentum part of the
problem. In particular, to find the solution at momentum $p_l$ we
integrate Eq. (\ref{EqPropagationSteadyEnLoss}) from $p_l$ to $p_{l+1}$
resulting in:
\begin{equation}
  \int\limits_{p_l}^{p_{l+1}}
  \frac{\partial \dot p \psi}{\partial p} dp
  =
  \left. \dot p \psi \right|_{p=p_{l+1}}
  -
  \left. \dot p \psi \right|_{p=p_l}
  =
  \int\limits_{p_l}^{p_{l+1}}
  \left(
  s(\vec{r}, p, t)
  +
  \nabla \cdot \mathcal{D} \nabla \psi
  -
  \frac{1}{\tau} \psi
  \right) dp
\end{equation}
Here the integral on the right hand side is now evaluated using the
trapezoidal rule, leading to a second order accurate
representation. With this we find after some rearrangement:
\begin{align}
  \label{EqSteadyEvolUgly}
  -\nabla \cdot \left . \mathcal{D} \nabla \psi \right|_{p=p_l}
  -
  \frac{2 \left.\dot p \psi\right|_{p=p_l}}{p_{l+1} - p_l} 
  + 
  \frac{\left. \psi \right|_{p=p_{l}}}{\tau}
  =&
  \\
  \nabla \cdot \left . \mathcal{D} \nabla \psi \right|_{p=p_{l+1}}
  -
  \frac{2 \left.\dot p \psi\right|_{p=p_{l+1}}}{p_{l+1} - p_l} 
  -&
  \frac{\left. \psi \right|_{p=p_{l+1}}}{\tau}
  +
  s(p_l, t)
  +
  s(p_{l+1}, t)
  \nonumber
\end{align}
where all terms containing $\left. \psi \right|_{p_l}$ are on the left
hand side of the equation. The right hand side with terms depending on
$\left. \psi \right|_{p_{l+1}}$ can, thus, be used as a source term,
when computing the solution at momentum $p_l$. Computing the solution
at the lower momentum using the one at higher momentum as a source
term reflects the fact that we are dealing with losses in momentum
space. Numerically speaking this means that we have advection from
high to low momenta. This also hints at the viable form for the
boundary conditions in momentum space: it is sufficient to prescribe
some value for the Cosmic Ray distribution function at the highest
momentum. For all physically meaningful propagation simulations we
will use the assumption that the distribution function is zero at the
highest energies.

From this discussion it is obvious that we can bring
Eq. (\ref{EqSteadyEvolUgly}) into a more convenient form. When
introducing the abbreviations:
\begin{equation}
  \Lambda
  =
  \frac{1}{\tau} -
  \frac{2 \left.  \dot p \right|_{p=p_{l}}}{p_{l+1} - p_l} 
\end{equation}
and
\begin{equation}
  S =
  \nabla \cdot \left . \mathcal{D} \nabla \psi \right|_{p=p_{l+1}}
  -
  \frac{2 \left.\dot p \psi\right|_{p=p_{l+1}}}{p_{l+1} - p_l} 
  -
  \frac{\left. \psi \right|_{p=p_{l+1}}}{\tau}
  +
  s(p_l, t)
  +
  s(p_{l+1}, t)
\end{equation}
we can rewrite Eq. (\ref{EqSteadyEvolUgly}) as:
\begin{equation}
  \label{EqSteadyEvolNice}
  -\nabla \cdot \left . \mathcal{D} \nabla \psi \right|_{p=p_l}
  + \Lambda \left. \psi \right|_{p=p_l}
  = S
\end{equation}
With this we arrived at a semi-discrete form of
Eq. (\ref{EqPropagationSteadyEnLoss}) where so far only the momentum space
is discretised.  Differential operators in configuration space are
then discretised using finite differences. For example the spatial
diffusion term can be written in Cartesian coordinates as:
\begin{equation}
  \label{EqDiffCart}
  \nabla \cdot \mathcal{D} \nabla \psi
  =
  \frac{\partial}{\partial x} \left(D_{xx}
  \frac{\partial}{\partial x} \psi\right)
  +
  \frac{\partial}{\partial y} \left(D_{yy}
  \frac{\partial}{\partial y} \psi\right)
  +
  \frac{\partial}{\partial z} \left(D_{zz}
  \frac{\partial}{\partial z} \psi\right)
\end{equation}
where we assumed the diffusion tensor to be diagonal. Any of these
three terms is now discretised in analogy to the example:
\begin{align}
  \label{EqDiff1DDisc}
  \frac{\partial}{\partial x} \left(D_{xx}
  \frac{\partial}{\partial x} \psi\right)
  =&
  \frac{D_{xx}(x_{i+\half})}{\Delta x^2}
  \psi(x_{i+1})
  +
  \frac{D_{xx}(x_{i-\half})}{\Delta x^2} \psi(x_{i-1})
  \nonumber\\
  &
  -
  \frac{D_{xx}(x_{i+\half}) + D_{xx}(x_{i-\half})}{\Delta x^2} \psi(x_{i})
  \nonumber
  \\
  =& A_i \psi(x_{i-1}) - B_i \psi(x_{i}) + C_i \psi(x_{i+1})
\end{align}
The diffusion tensor can also contain off-diagonal elements $D_{xy}$,
$D_{xz}$ or $D_{yz}$, but needs to be symmetric in any case. In the
presence of off-diagonal elements the discrete version of
Eq. \eqref{EqDiffCart} has to be extended by terms of the form:
\begin{align}
  \frac{\partial}{\partial x}
  \left(D_{xy} \frac{\partial}{\partial y} \psi\right)
  &+
  \frac{\partial}{\partial y}
  \left(D_{xy} \frac{\partial}{\partial x} \psi\right)
  \\
  =&
  \frac{D_{xy}(x_{i+1}, y_{j}) + D_{xy}(x_{i}, y_{j+1})}{4\Delta x \Delta y}
  \psi(x_{i+1}, y_{j+1})
  \nonumber\\
  &-
  \frac{D_{xy}(x_{i+1}, y_{j}) + D_{xy}(x_{i}, y_{j-1})}{4\Delta x \Delta y}
  \psi(x_{i+1}, y_{j-1})
  \nonumber\\
  &-
  \frac{D_{xy}(x_{i-1}, y_{j}) + D_{xy}(x_{i}, y_{j+1})}{4\Delta x \Delta y}
  \psi(x_{i-1}, y_{j+1})
  \nonumber\\
  &+
  \frac{D_{xy}(x_{i-1}, y_{j}) + D_{xy}(x_{i}, y_{j-1})}{4\Delta x \Delta y}
  \psi(x_{i-1}, y_{j-1})
  \nonumber
\end{align}
with analogous expressions for the other possible entries.

With this we have a fully discrete form of the steady state transport
equation for Cosmic Rays. The resulting discretisation yields a
coupled system of equations for $\psi(x_i,y_j,z_k, p_l)$.  In this
case the values of $\psi$ at the different grid points are coupled due
to the fact that the discretisation of the diffusion operator in
Eq.(\ref{EqSteadyEvolNice}) at point $(x_i,y_j,z_k)$ yields a function
of $\psi$ at $(x_i,y_j,z_k)$ and at all neighbouring spatial
grid-points as well. Insofar we have to deal with an implicit
evolution scheme in momentum space. While there might be no efficient
direct solver for the corresponding matrix equation available there
are still several methods, which can solve such systems
efficiently. For all models computed in this paper we used an a
Gauss-Seidel multigrid solver.

For this solver we explicitly made use of the fact that in momentum
space the solution at momentum $p_l$ can be computed using the one at
momentum $p_{l+1}$ as input. This is possible due to the fact that we
are exclusively dealing with momentum losses in the transport
equation. Therefore, we are confronted with a coupling of the
grid-points in the spatial domain only, accordingly requiring a solver
for a three-dimensional instead of a four-dimensional problem. In
principle this approach can also be extended to the case where the
Cosmic Rays are subject to energy gains over the whole computational
domain instead of energy losses. While this may be the case for
re-acceleration models, they pose the additional problem that also a
diffusion term in momentum space is present, also leading to a
coupling to the adjacent cells in momentum space. This case will be
addressed in Sec. \ref{SecReaccScheme}.

\subsubsection{Numerical solution of the 3D matrix equation}
\label{SecMatrixSolve}
As discussed in the previous section, after discretisation we are
facing a coupled system of algebraic equations. This system can be
rewritten in matrix form. This matrix equation would for one spatial
dimension be a tridiagonal matrix. For such a case there are numerical
methods for the direct inversion of the matrix available where the
numerical cost is of order $\mathcal{O}(N)$. For three-dimensional
problems, however, there are no such methods available
anymore. Therefore we will in this case rely on an iterative method to
solve the system of equations.

There are several such iterative relaxation methods available like,
e.g. Jacobi or Gauss-Seidel just to name a few. These methods,
however, suffer from the fact that the relaxation is most efficient
for short wavelengths, while long wavelength relaxation takes a large
amount of iterations. To remedy this shortcoming the multigrid method
was introduced, which applies grids of different resolution to allow
an efficient relaxation at all wavelengths of a problem. It turns out
that multigrid methods can be set up that are also of order
$\mathcal{O}(N)$ with regard to numerical expense.

Here we are using Gauss-Seidel relaxation in a multigrid method. We
included different variations of this method in the code. The
difference is in the order in which the solutions is applied to the
different grid-points. Here we are using either so called red-black
Gauss-Seidel or an alternating plane Gauss-Seidel version \citep[see,
  e.g.,][for further details]{TrottenbergEtAlBook2001,
  TholeTrottenberg86AMaC19_333}. The former of these is faster
regarding a single iteration but the latter turns out to need fewer
iterations for anisotropic problems like, e.g., strongly anisotropic
source distributions. With this we can now discuss the implementation
of the re-acceleration scheme.

\subsection{Re-acceleration scheme}
\label{SecReaccScheme}
In the presence of re-acceleration we are confronted with a coupled
system of equations in all four dimensions. This requires a different
approach to the discretisation in momentum space and also some
extensions to the matrix solver.

\subsubsection{Discretisation}
 For the re-acceleration scheme we use the same discretisation for the
 spatial derivatives as was discussed for the energy-loss scheme. For
 the momentum derivatives, in contrast to the approach described under
 \ref{SecDiscMomLoss}, we will also use a finite difference
 discretisation. We note that the re-acceleration term can be
 expressed as a combination of a momentum diffusion and an advection
 term in momentum space that represents an energy gain:
\begin{equation}
\label{EqFermiII}
 \frac{\partial}{\partial p}
  \left(
  p^2 D_{pp} \frac{\partial}{\partial p} \frac{\psi}{p^2}
  \right)
  =
  \frac{\partial}{\partial p}
  \left(
  D_{pp}
  \frac{\partial \psi}{\partial p}
  \right)
  +
  \frac{\partial}{\partial p}
  \left(
  - \frac{2 D_{pp}}{p} \psi
  \right).
\end{equation}
With this we find for the momentum advection velocity $u_p$:
\begin{equation}
  u_p =
  \frac{2 D_{pp}}{p}
  +
  \dot p
  -
  \frac{p}{3} \left(\nabla \cdot \vec{u}\right).
\end{equation}
This expression shows that the momentum advection velocity may have a
different sign at different energies or positions, also
reflected in the corresponding discretisation:
\begin{equation}
  \label{EqMomUpwind}
  \frac{\partial u_p \psi}{\partial p} =
  \left\{
  \begin{array}{cl}
    \displaystyle
    \frac{u_{p;l+1}\psi_{l+1} -u_{p;l} \psi_l}{p_{l+1} - p_l}
    &\qquad\text{if } u_{p;l} < 0
    \\
    \displaystyle
    \frac{u_{p;l}\psi_l -u_{p;l-1} \psi_{l-1}}{p_l - p_{l-1}}
    &\qquad\text{else}
  \end{array}
  \right .
\end{equation}
which is the same as used in the \textsc{Galprop} code.  At the same
time we discretise the diffusion term via:
\begin{align}
  \frac{\partial}{\partial p}
  \left(
  D_{pp}
  \frac{\partial \psi}{\partial p}
  \right)
  &=
  \frac{1}{p_{l+\half} - p_{l-\half}} \left(
  \frac{ D_{pp}(p_{l+\half})}{p_{l+1} - p_{l}}
  \psi_{l+1}
  \right.
  \\
  \nonumber
  &\qquad\left.
  -
  \left(
  \frac{ D_{pp}(p_{l+\half})}{p_{l+1} - p_{l}} +
  \frac{D_{pp}(p_{l-\half})}{p_{l} - p_{l-1}}
  \right)
  \psi_l
  +
  \frac{D_{pp}(p_{l-\half})}{p_{l} - p_{l-1}} \psi_{l-1}
  \right)
\end{align}
This shows that the discretisation of the transport equation at $(x_i,
y_j, z_k, p_l)$ now also depends on the value of $\psi$ at the two
neighbouring points in momentum space. Thus, in this case we are
dealing with a four-dimensional matrix problem.

\newpage
\subsubsection{Numerical solution of the 4D matrix equation}
To solve the four-dimensional coupled system of algebraic equations we
apply a four-dimensional extension of the multigrid solver discussed
in Sec. \ref{SecMatrixSolve}. Here we found that a four-dimensional
red-black Gauss-Seidel iteration does not yields satisfactorily
convergence rates.  In this red-black Gauss-Seidel approach the
solution at a single grid-point is computed using the values of the
surrounding grid-points in all four dimensions. In contrast to that it
turns out to be significantly more efficient to compute the solution
collectively for all spatial grid-points at the same momentum. For
this collective solution of the spatial problem we use a few
iterations of the three-dimensional solution scheme presented in
Sec. \ref{SecMatrixSolve}.  At each multigrid level of the
four-dimensional multigrid solver this approach is applied at all
grid-points in momentum space. In this we alternate between the
solution at all even and all odd grid-points in momentum. This is due
to the chosen discretisation where, e.g., the solution at an even
grid-point in momentum only depends directly on the solution at the
surrounding odd grid-points.

Corresponding tests that validate the capabilities of these solvers
are given in Section \ref{SecTests}. We found that both the
three-dimensional and the four-dimensional multigrid solver yield a
solution after $\sim$10 iterations. With this the steady state method
is significantly faster than any method that attempts to find a steady
state solution using a time integration scheme. Even the time
integration scheme used within the \textsc{Galprop} code that uses
ever decreasing timesteps
\citep[see][]{StrongMoskalenko1998APJ509_212} is slower than the
present steady state algorithm. This is of particular interest as
\textsc{Galprop} aims at finding the steady state solution of the
propagation problem.

Having found a suitable steady state solution is a promising starting
point for a time dependant computation with, e.g, time variable
sources. This is also handled by an adapted time integration scheme.

\subsection{The time integration scheme}
When time dependant parameters (like, e.g., a variable source
distribution) are present we have to use the evolution equation
Eq. \eqref{EqPropagation} instead of
Eq. \eqref{EqPropagationSteady}. In this case the only difference is
the presence of the time derivative. This, however, poses no problem
in using the same momentum space discretisation as for the previously
discussed steady state solvers.

Here we begin the discussion again with the energy-loss scheme. At
this point we just repeat the corresponding derivation including the
rate of change in time. In applying the momentum space integration to
the dynamical problem we find as a first step:
\begin{align}
  \label{EqDynEvolUgly}
  \left.\frac{\partial \psi}{\partial t} \right|_{p=p_l}
  -\nabla \cdot \left . \mathcal{D} \nabla \psi \right|_{p=p_l}
  -
  \frac{2 \left.\dot p \psi\right|_{p=p_l}}{p_{l+1} - p_l} 
  + 
  \frac{\left. \psi \right|_{p=p_{l}}}{\tau}
  =&
  \\
  -\left.\frac{\partial \psi}{\partial t} \right|_{p=p_{l+1}} +
  \nabla \cdot \left . \mathcal{D} \nabla \psi \right|_{p=p_{l+1}}
  -
  \frac{2 \left.\dot p \psi\right|_{p=p_{l+1}}}{p_{l+1} - p_l} 
  -&
  \frac{\left. \psi \right|_{p=p_{l+1}}}{\tau}
  \nonumber\\
  +
  s(p_l, t)
  +
  s(p_{l+1}, t)
  &
\nonumber
\end{align}
This is so far only discretised in momentum space. As a next step we
therefore perform a similar integration in time. While this would in
principle allow different timestep sizes, using such would be highly
unusual because commonly the timestep size is determined either from
energy loss timescales or particle decay timescales. These, however,
usually do not change with time. Using the same second order integral
as for the momentum discretisation we find:
\begin{align}
  \label{EqDynEvolExtreme}
  &2\left.\frac{\psi^{n+1} - \psi^n}{\Delta t} \right|_{p=p_l}
  -
  \nabla \cdot \left . \mathcal{D} \nabla \psi^{n+1} \right|_{p=p_l}
  -
  \nabla \cdot \left . \mathcal{D} \nabla \psi^{n} \right|_{p=p_l}
  \nonumber\\
  &\qquad-
  \frac{2 \left.\dot p \psi^{n+1}\right|_{p=p_l}}{p_{l+1} - p_l} 
  -
  \frac{2 \left.\dot p \psi^n\right|_{p=p_l}}{p_{l+1} - p_l} 
  + 
  \frac{\left. \psi^{n+1} + \psi^n \right|_{p=p_{l}}}{\tau}
  =
  \nonumber\\
  &-2\left.\frac{\psi^{n+1} - \psi^n}{\Delta t} \right|_{p=p_{l+1}}
  +
  \nabla \cdot \left . \mathcal{D} \nabla \psi^{n+1} \right|_{p=p_{l+1}}
  +
  \nabla \cdot \left . \mathcal{D} \nabla \psi^n \right|_{p=p_{l+1}}
  \\
  &\qquad-
  \frac{2 \left.\dot p \psi^{n+1}\right|_{p=p_{l+1}}}{p_{l+1} - p_l} 
  -
  \frac{2 \left.\dot p \psi^{n}\right|_{p=p_{l+1}}}{p_{l+1} - p_l} 
  -
  \frac{\left. \psi^{n+1} + \psi^n \right|_{p=p_{l+1}}}{\tau}
  \nonumber\\
  &\qquad+
  s(p_l, t^{n+1})
  +
  s(p_l, t^{n})
  +
  s(p_{l+1}, t^{n+1})
  +
  s(p_{l+1}, t^n)
  \nonumber
\end{align}
where $n$ indicates the timestep. Here $t^n$ is the timestep for which
all quantities are known and the quantities at $t^{n+1}$ are to be
computed. From this one can actually find a very similar evolution
scheme as was used for the solution of the steady state problem. Due
to the fact that we only consider energy \emph{loss} processes in this
paper quantities at momentum $p_l$ depend on such at momentum
$p_{l+1}$ which was already used for the steady state problem. Then
the procedure following from the above discretisation is as follows:
after having computed $\psi^n$ we start the computation of
$\psi^{n+1}$. For this we apply boundary conditions at the highest
momentum. For $N$ gridpoints in the computational domain of momentum
space (with the range $p = p_0 \dots p_{N-1}$) we then use the
boundary condition at $p_{N}$. Then we can compute the solution at
$t^{n+1}, p_{n-1}$ from the values at the previous timestep and the
boundary values. From the resulting
$\left.\psi^{n+1}\right|_{p=p_{N-1}}$ we can then compute
$\left.\psi^{n+1}\right|_{p=p_{N-2}}$ and so on. That means we can
easily rearrange Eq. \eqref{EqDynEvolExtreme} in a way that all
unknown quantities are on the left hand side. In particular we find:
\begin{align}
  \label{EqDynEvolExtremeSort}
  -
  \nabla \cdot &\left . \mathcal{D} \nabla \psi^{n+1} \right|_{p=p_l}
  -
  \frac{2 \left.\dot p \psi^{n+1}\right|_{p=p_l}}{p_{l+1} - p_l} 
  + 
  \left( \frac{1}{\tau} + \frac{2}{\Delta t}\right)
  \left. \psi^{n+1}\right|_{p=p_{l}}
  =
  \\
  =&
  +
  \nabla \cdot \left . \mathcal{D} \nabla \psi^{n+1} \right|_{p=p_{l+1}}
  -
  \frac{2 \left.\dot p \psi^{n+1}\right|_{p=p_{l+1}}}{p_{l+1} - p_l} 
  -
  \left(\frac{1}{\tau} + \frac{2}{\Delta t}\right)
  \left. \psi^{n+1} \right|_{p=p_{l+1}}
  \nonumber\\
  &+
  \nabla \cdot \left . \mathcal{D} \nabla \psi^{n} \right|_{p=p_l}
  +
  \frac{2 \left.\dot p \psi^n\right|_{p=p_l}}{p_{l+1} - p_l} 
  -
  \left(\frac{1}{\tau} - \frac{2}{\Delta t}\right)
  \left.\psi^n \right|_{p=p_{l}}
  \nonumber\\
  &+
  \nabla \cdot \left . \mathcal{D} \nabla \psi^n \right|_{p=p_{l+1}}
  -
  \frac{2 \left.\dot p \psi^{n}\right|_{p=p_{l+1}}}{p_{l+1} - p_l} 
  -
  \left(\frac{1}{\tau} - \frac{2}{\Delta t}\right)
  \left.\psi^n \right|_{p=p_{l+1}}
  \nonumber\\
  &+
  s(p_l, t^{n+1})
  +
  s(p_l, t^{n})
  +
  s(p_{l+1}, t^{n+1})
  +
  s(p_{l+1}, t^n)
  \nonumber
\end{align}
This equation can actually be brought into the exact same form as the
evolution equation for the steady state momentum-loss problem
Eq. \eqref{EqSteadyEvolNice}. The above can be written as:
\begin{equation}
  \label{EqEvolNice}
  -\nabla \cdot \left . \mathcal{D} \nabla \psi^{n+1} \right|_{p=p_l}
  + \tilde \Lambda \left. \psi^{n+1} \right|_{p=p_l}
  = \tilde S
\end{equation}
by using the definitions:
\begin{equation}
  \tilde \Lambda
  =
  \left( \frac{1}{\tau} + \frac{2}{\Delta t}\right) -
  \frac{2 \left.\dot p\right|_{p=p_l}}{p_{l+1} - p_l} 
\end{equation}
and
\begin{align}
  \tilde S 
  =&
  \nabla \cdot \left . \mathcal{D} \nabla \psi^{n+1} \right|_{p=p_{l+1}}
  -
  \frac{2 \left.\dot p \psi^{n+1}\right|_{p=p_{l+1}}}{p_{l+1} - p_l} 
  -
  \left(\frac{1}{\tau} + \frac{2}{\Delta t}\right)
  \left. \psi^{n+1} \right|_{p=p_{l+1}}
  \\
  &+
  \nabla \cdot \left . \mathcal{D} \nabla \psi^{n} \right|_{p=p_l}
  +
  \frac{2 \left.\dot p \psi^n\right|_{p=p_l}}{p_{l+1} - p_l} 
  -
  \left(\frac{1}{\tau} - \frac{2}{\Delta t}\right)
  \left.\psi^n \right|_{p=p_{l}}
  \nonumber\\
  &+
  \nabla \cdot \left . \mathcal{D} \nabla \psi^n \right|_{p=p_{l+1}}
  -
  \frac{2 \left.\dot p \psi^{n}\right|_{p=p_{l+1}}}{p_{l+1} - p_l} 
  -
  \left(\frac{1}{\tau} - \frac{2}{\Delta t}\right)
  \left.\psi^n \right|_{p=p_{l+1}}
  \nonumber\\
  &+
  s(p_l, t^{n+1})
  +
  s(p_l, t^{n})
  +
  s(p_{l+1}, t^{n+1})
  +
  s(p_{l+1}, t^n)
  \nonumber
\end{align}
This equation can, again, be solved by a discretisation of the
diffusion operator and subsequent application of the numerical matrix
solver.

For the re-acceleration solver we can use a similar treatment. In this
case we just need to add the time-derivative to the discrete form of
the transport equation. Then by an integral over one timestep as
discussed above we actually come up with the Crank-Nicolson
discretisation also used in \textsc{Galprop}, where all terms at time
$t^n$ are used as source terms and we are solving for
$\psi^{n+1}$. The resulting equation then reads:
\begin{align}
  &-
  \nabla \cdot\left( \mathcal{D} \nabla \psi^{n+1}\right)
  + 
  \nabla \cdot \left(\vec{u} \psi^{n+1}\right)
  -
  \frac{\partial}{\partial p}
  \left(
  p^2 D_{pp}\frac{\partial}{\partial p}\frac{\psi^{n+1}}{p^2}
  \right)
  \\
  \nonumber
  &\qquad+
  \frac{\partial}{\partial p}
  \left(
  \dot p \psi^{n+1} - \frac{p}{3}\left(\nabla\cdot\vec{u}\right) \psi^{n+1}
  \right)
  + 
  \frac{1}{\tau} \psi^{n+1}
  +
  \frac{2}{\Delta t} \psi^{n+1}
  \\
  \nonumber
  &=
  s(\vec{r}, p, t^n)
  +
  s(\vec{r}, p, t^{n+1})
  +
  \nabla \cdot\left( \mathcal{D} \nabla \psi^{n}\right)
  -
  \nabla \cdot \left(\vec{u} \psi^{n}\right)
  +
  \frac{\partial}{\partial p}
  \left(
  p^2 D_{pp}\frac{\partial}{\partial p}\frac{\psi^{n}}{p^2}
  \right)
  \\
  \nonumber
  &\qquad-
  \frac{\partial}{\partial p}
  \left(
  \dot p \psi^{n} - \frac{p}{3}\left(\nabla\cdot\vec{u}\right) \psi^{n}
  \right)
  - 
  \frac{1}{\tau} \psi^{n}
  +
  \frac{2}{\Delta t} \psi^{n}
\end{align}
By combining all terms on the right-hand side into a single source
term, and using the substitution:
\begin{equation}
  \frac{1}{\tilde \tau} \psi^{n+1}
  =
  \frac{1}{\tau} \psi^{n+1}
  +
  \frac{2}{\Delta t} \psi^{n+1}
\end{equation}
we end up with an equation of the same form as
Eq. \eqref{EqPropagationSteady}. This means that we find an equation
of the same form as for the steady state problem in both cases.
Accordingly, the same numerical solver can be applied to all
problems.

\subsection{Parameters and improved nuclear reaction network}
The portrayed numerical solver is embedded into a computational
framework that supplies all relevant parameters at runtime. These
parameters are in particular the spatial diffusion tensor, the momentum
loss rates, the source distribution and the timescales for radioactive
and fragmentation decays (see Eq. \eqref{EqPropagation}), all of which
might vary with spatial position and energy. Furthermore they depend on
physical quantities like, e.g., the gas distribution, nuclear cross
sections and intensities of the interstellar radiation fields for
which we adopt here the same distributions as in
\textsc{Galprop}. Additionally we allowed for a spatial variation of
the components of the diffusion tensor.


The solver is applied to all nuclei individually which are coupled
through radiative decay and fragmentation, where the latter results
from an interaction with the interstellar medium. This means that
heavier nuclei can decay into lighter ones, and, consequently, serve
as a secondary source for the latter.  Therefore the solver is first
applied to the heaviest particle with mass-number $A$ and charge
$Z$. In the next step the solver is applied to the isotope with $A-1$
until the solution for all isotopes of the current element is
found. Then the code continues with the element with $Z-1$ in the same
fashion. In this way the transport problem for most of those species
that can act as a secondary source for the species currently under
investigation was already treated earlier.

There are, however, a few exception where species that appear later in
the structure of the network can decay into an earlier one. The most
commonly known example is probably the beta-decay of
$\nucl{10}{4}{Be}$ into $\nucl{10}{5}{B}$. \textsc{Galprop} deals with
this problem by iterating several times over the full nuclear reaction
network, which increases the numerical cost of the code linearly with
the number of iterations. We are using a different approach: First the
code determines those elements, which have a secondary source partner
earlier in the network. During the computation the solver then just
iterates over the range between these two coupled elements instead of
an iteration of the whole network. Using this numerical scheme is not
only cheaper due to the fact that we avoid the re-computation of the
complete reaction network, also the numerical solution of species
solved repeatedly in the network iterations becomes significantly
faster. This is due to the fact that the solution at the previous
iteration is used as an input, which, however, does not deviate
significantly from the final solution anymore. Therefore the
\textsc{Picard} code is even more efficient when the nuclear reaction
network is fully explored.

\section{Validation of the scheme}
\label{SecTests}
In this section we will use several simplified forms of the evolution
equation Eq. \eqref{EqPropagation} to proof the correctness of the
different numerical schemes. All the following tests have an
analytical solution, thus, allowing a direct evaluation of the quality
of the numerical scheme.

\subsection{Energy loss test}
\label{SecTestMom}
This first test will be used to verify the convergence properties of
the solution in momentum space for the energy-loss scheme. For that we
will only take losses and source terms into account and assume that we
have a steady state problem. From
Eq. (\ref{EqPropagationSteadyEnLoss}) we see that we are dealing with
the equation:
\begin{equation}
  \label{EqMomOnly}
  \frac{d \dot p \psi}{d p} = s(p)
\end{equation}
At this point we need a specific form for the loss term $\dot p$ and
for the source term $s(p)$. Here we will just use different power-laws
for each. That is, we will use:
\begin{equation}
  \dot p = - a \left(\frac{p}{p_0}\right)^n;
  \qquad
  s(p) = s_0 \left(\frac{p}{p_0}\right)^{-s}
\end{equation}
where $p_0$ is a reference momentum. This differential equation can
easily be solved by standard methods, where the corresponding
analytical solution is:
\begin{equation}
  \psi
  =
  C \left(\frac{p}{p_0}\right)^{-n}
  +
  \frac{1}{s-1} \frac{s_0 p_0}{a} \left(\frac{p}{p_0}\right)^{1- (s+n)}
\end{equation}
The different constants are chosen in a way that both terms contribute
significantly to the resulting spectrum. First we use the following
values for the power-indices: $n={1.5}$ and $s=2.2$. Then we have:
\begin{equation}
  \psi
  =
  C \left(\frac{p}{p_0}\right)^{-1.5}
  +
  \frac{5}{6} \frac{s_0 p_0}{a} \left(\frac{p}{p_0}\right)^{-2.7}
\end{equation}
The different constants used within this test problem might not be
physically meaningful, but they will still serve as good examples for
testing the solver. With the momentum given in units of $p_0 = 1$ MeV,
we use $s_0 = 10^6 a/p_0$ in order to have a spectral break at about
$10^5$ MeV. The test is run on a grid similar to what is eventually
used in Cosmic Ray transport problems. That is we use a kinetic energy
between 10 MeV and $10^9$ MeV for our test, subdivided into $N$
logarithmically spaced cells.

We use this as a test for the momentum discretisation of both the
steady-state and the time-dependant version of the solver. For the
former test we use Eq. (\ref{EqSteadyEvolUgly}) where we neglect
spatial diffusion and catastrophic losses. With this we find the
algebraic equation:
\begin{equation}
  \left . \psi \right|_{p=p_{l}}
  =
  \frac{\left . \dot p \right|_{p=p_{l+1}}}
       {\left . \dot p \right|_{p=p_{l}}} 
       \left . \psi \right|_{p=p_{l+1}} 
  -
  \frac{1}{2\left . \dot p \right|_{p=p_{l}}}\left(s(p_{l}) + s(p_{l+1})\right)\left(p_{l+1} -p_l\right)
\end{equation}
In this case we prescribe the analytical solution at $p = p_N$ and
compute the values at lower momenta successively.

For the time-dependant version we use a very similar procedure. In
this case we keep all relevant terms from
Eq. \eqref{EqDynEvolExtremeSort}, which results in:
\begin{align}
  \left. \psi^{n+1}\right|_{p=p_{l}}
  &=
  \left(
  \frac{\Delta t \left.\dot p\right|_{p=p_{l+1}} + \Delta p}
       {\Delta t \left.\dot p \right|_{p=p_l} - \Delta p}
  \right)
  \left. \psi^{n+1} \right|_{p=p_{l+1}}
  \nonumber\\
  &-
  \left(
  \frac{\Delta t \left.\dot p \right|_{p=p_l} + \Delta p}
       {\Delta t \left.\dot p \right|_{p=p_l} - \Delta p}
  \right)
  \left.\psi^n \right|_{p=p_{l}}
  +
  \left(
  \frac{\Delta t \left.\dot p \right|_{p=p_{l+1}} - \Delta p}
       {\Delta t \left.\dot p \right|_{p=p_l} - \Delta p}
  \right)
  \left.\psi^n \right|_{p=p_{l+1}}
  \nonumber\\
  &+\frac{\Delta t \Delta p}
      {2\left(\Delta t \left.\dot p \right|_{p=p_l} - \Delta p\right)}
  \left(
  s(p_l, t^{n+1})
  +
  s(p_l, t^{n})
  \right.
  \nonumber\\
  &\qquad\qquad\qquad\qquad\qquad
  \left.
  +
  s(p_{l+1}, t^{n+1})
  +
  s(p_{l+1}, t^n)
  \right)
\end{align}
where $\Delta p = p_{l+1} - p_l$. In this case the solution depends on
the solution at the previous timestep and the one at the current time
but at momentum $p_{l+1}$. Apart from that we also prescribe the
analytical solution at $p=p_N$.

\begin{figure}
  \begin{center}
    \setlength{\unitlength}{0.008cm}
    \begin{picture}(1100,862)(-100,-100)
      \put(330,-70){Number of gridpoints}%
      \put(-70,360){\rotatebox{90}
        {L2 error}}%
      \includegraphics[width=1000\unitlength]{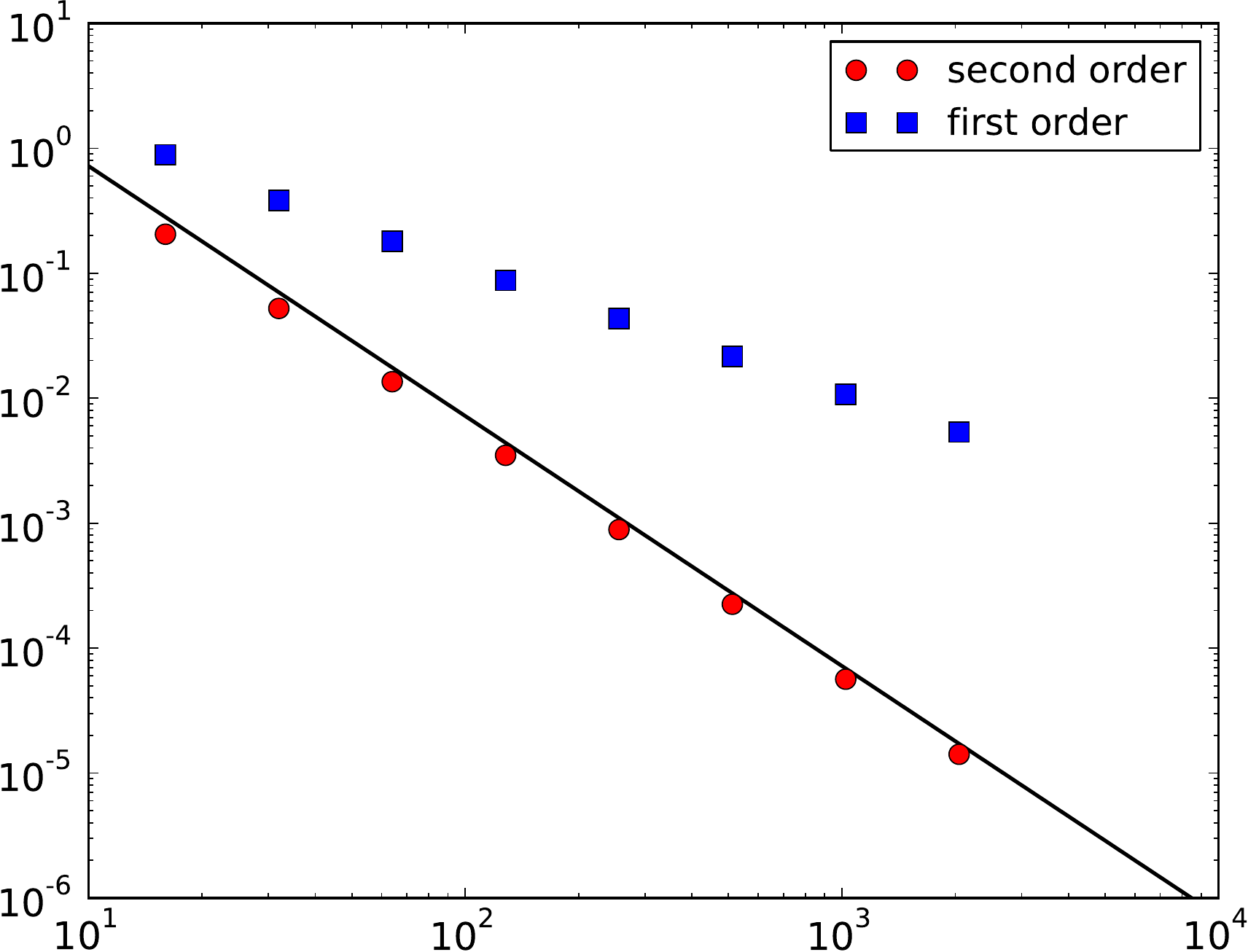}
    \end{picture}
  \end{center}
  \caption{\label{FigMomTestOrder}Relative L2 error for the momentum space
    test. Here we show the errors as a function of the number of
    gridpoints for the steady state solver (red circles for the second
    order and blue squares for the first order version). Results for
    the time dependant version are identical. For comparison the black
    line indicates a slope of -2.}
\end{figure}

For a solution via the steady state method we find that the results
are in agreement with the analytical solution and that the method is
indeed of second order in momentum as can be seen from
Fig. \ref{FigMomTestOrder}, where the relative l2 error is shown as a
function of the number of gridcells. This error is computed as:
\begin{equation}
  \epsilon_{l2,rel}
  =
  \sqrt{
    \frac{\sum_{l=0}^{N-1} \left(\psi^{num}(p_l) - \psi^{ana}(p_l)\right)^2}
  {\sum_{l=0}^{N-1}\left(\psi^{ana}(p_l)\right)^2}
  }
\end{equation}
where $\psi^{num}$ is the solution computed by the numerical scheme
and $\psi^{ana}$ is the correct, analytical result.

Additionally we show results for a first order version of the steady
state solver. Apparently, for a resolution of 100 gridpoints an error
in the order of \mbox{10 \%} can be expected in that case. This shows
that for a reasonable number of gridpoints in momentum space a second
order solver is recommended, for which the error at the given
resolution is well below the percent level. Note that we are dealing
with a four-dimensional problem, so excessively high numbers of
gridpoints are not feasible due to memory constraints.

For the test of the time-dependant solver the problem is initialised
with $\psi(p) = 1$ everywhere apart from the upper boundary. Then the
solver is applied successively until a steady state solution is
reached. Again, the result is consistent with the analytical solution
and the code is of second order as expected. Actually we found that
when the steady state solution is reached with the time-dependant
solver, the result is identical to the one found using the
steady-state solver.

One important aspect, however, became obvious from this test. When
using time integration to find a steady state solution the number of
timesteps necessary to reach the solution can be fairly high. This is
due to the fact that the loss term is represented by a power-law in
momentum space. This means that the problem relaxes quickly at highest
energies, whereas relaxation at low energies takes much
longer. Actually the correct solution is slowly propagated down to the
smallest energies. This problem is somewhat remedied by using the
logarithmic grid, but still the number of timesteps necessary to reach
a steady state solution can be fairly large. This has to be kept in
mind, whenever trying to reach a steady state using a time integration
scheme to solve the energy loss problem in case of significant changes
of the loss timescale with energy. In particular the direct solution
via the steady state solver is much more efficient in that regard.

\subsection{Diffusion test}
\label{SecTestDiff}
Having shown that the solution for the momentum space problem is
accurate we now need to verify the solver for the spatial diffusion
problem. In particular we need to verify that the solver can handle a
problem of the form:
\begin{equation}
  \label{EqDiffOnly}
  - \nabla \cdot \mathcal{D} \nabla \psi = s(\vec{r})
\end{equation}
Together with the boundary conditions $\psi_r(x=\pm R, y=\pm R, z=\pm
H) = 0$, typically used for Galactic propagation models, it is rather
simple to come up with a test problem that has an analytical
solution. Here we will just prescribe an analytical solution as:
\begin{equation}
  \psi_{ana} = \cos{\frac{x \pi}{2R}}\cos{\frac{y \pi}{2R}}\cos{\frac{z \pi}{2H}}
\end{equation}
where the numerical domain extents from $-R$ to $R$ in the $x$- and
$y$-direction and from $-H$ to $H$ in the $z$-direction. For this we
then find the corresponding source-function that fulfils
Eq. (\ref{EqDiffOnly}) as:
\begin{equation}
  \label{EqDiffTestSrc1}
  s(\vec{r})
  =
  \frac{\pi^2}{4}
  \left(
  \frac{D_{xx} + D_{yy}}{R^2} + \frac{D_{zz}}{H^2}
  \right)
  \psi_{ana}
\end{equation}
We test the diffusion solver by using this source function within
Eq. (\ref{EqDiffOnly}) where we initialise $\psi$ as zero
everywhere. The entries of the diffusion tensor are spatially constant
with: $D_{xx} = D_{yy} = 10^{24}$ m$^2$ s$^{-1}$ and $D_{zz} =
10^{26}$ m$^2$ s$^{-1}$.  This test might appear simplified, but at a
closer look the source function is strongly variable and the problem
also shows a significant anisotropy in the $z$-direction.

Tests of this problem show that our numerical solver can very
efficiently handle this specific problem. Already for a spatial
resolution of 33 gridcells we find a deviation from the analytical
solution of $<0.1$\%. By studying different resolutions like for the
previous test we can verify the second order accuracy of the
method. The present problem turns out to be sufficiently anisotropic
to demand an alternating plane relaxation technique (see
Sec. \ref{SecMatrixSolve}), for which it converges in just a few
iterations.

As another test we also investigate the case with spatially variable
entries of the diffusion tensor. In particular we use $D_{xx} =
x^2$, $D_{yy} = z^2$ and $D_{zz} = y^2$. This problem is obviously
significantly anisotropic. Using the source term:
\begin{equation}
  s(\vec{r})
  =
  \frac{x\pi}{R} \sin{\frac{x\pi}{2R}}
  \cos{\frac{y\pi}{2R}} \cos{\frac{z\pi}{2H}}
  +
  \frac{\pi^2}{4}
  \left(
  \frac{D_{xx} + D_{yy}}{R^2} + \frac{D_{zz}}{H^2}
  \right)
  \psi_{ana}
\end{equation}
actually corresponds to the same analytical solution as before. Also
for this case we observe the expected second order convergence. Here
the deviation from the analytical solution is somewhat larger with 0.18
\% at a spatial resolution of 33 gridpoints in each dimension. Both
tests have been verified up to 129 gridpoints in each dimension being
equivalent to a spatial resolution of about 0.3 kpc in the $x$- and
$y$-directions and 0.047 kpc in the $z$-direction where they show
deviations from the analytical solution of $\le10^{-4}$ for the former
of these tests. This also shows that adequate handling of these high
resolutions is indeed possible with the new method.



\subsubsection{Arbitrary direction diffusion}
The most general diffusion test we applied is one including non-zero
off-diagonal elements. For this an alternating plane relaxation
technique turns out to be required, too. We assume that diffusion is
dominant along a preferred direction and reduced in perpendicular
directions. In this case this is the cylindrical
$\phi$-direction. That leads to the following form for the diffusion
tensor:

\begin{equation}
  \mathcal{D}
  =
  \left(
  \begin{array}{ccc}
    D_{xx}
    &
    D_{xy}
    &
    0
    \\
    D_{xy}
    &
    D_{yy}
    &
    0
    \\
    0
    &
    0
    &
    D_{\perp}
  \end{array}
  \right)
\end{equation}
where
\begin{align}
  D_{xx} &= D_{\parallel} \sin^2 \phi + D_{\perp} \cos^2 \phi;
  \qquad
  D_{yy} = D_{\parallel} \cos^2 \phi + D_{\perp} \sin^2 \phi;
  \\
  D_{xy} &= (D_{\perp} - D_{\parallel}) \sin \phi \cos \phi.
  \nonumber
\end{align}
For this case we assume that the diffusion tensor in the local frame
relative to the $\phi$ dimension features constant entries, i.e., we
use $D_{\parallel} = 10^{24}$\,m$^2$\,s$^{-1}$ and $D_{\perp} =
10^{23}$\,m$^2$\,s$^{-1}$. Using the same analytical solution as
before we find that the source-function needs to fulfill:
\begin{align}
  s(\vec{r})
  =&
  \frac{\pi^2}{4}\left(\frac{D_{xx} + D_{yy}}{R^2}
  +
  \frac{D_{zz}}{H^2}\right) \psi
  -
  D_{xy}\frac{\pi^2}{2R^2}
  \sin\frac{x \pi}{2R} \sin\frac{y \pi}{2 R} \cos\frac{z \pi}{2 H}
  \\
  &+
  \left(
  \frac{\partial D_{xx}}{\partial x} +
  \frac{\partial D_{xy}}{\partial y}
  \right)
  \frac{\pi}{2R} \sin\frac{x \pi}{2R} \cos\frac{y \pi}{2R} \cos\frac{z\pi}{2H}
  \nonumber\\
  &+
  \left(
  \frac{\partial D_{yy}}{\partial y} +
  \frac{\partial D_{xy}}{\partial x}
  \right)
  \frac{\pi}{2R} \cos\frac{x \pi}{2R} \sin\frac{y \pi}{2R} \cos\frac{z \pi}{2H},
  \nonumber
\end{align}
where we implicitly used that $D_{zz} = D_{\perp} = const.$ Like for
the previous tests we computed solutions up to a resolution of 129
gridpoints in each dimension. The deviations from the analytical
solution are similar to the previous tests, where we find in this case
an error of 0.13\% with 33 gridpoints and a second order reduction of
the error with incresing spatial resolution. This shows that the
resulting deviation stems from the discretisation only. The chosen
numerical scheme for the solution of the diffusion problem solves the
discrete problem exactly.


\subsection{Overall analytical test}
\label{SecTestOverall}
We conclude the discussion of analytical tests with one that includes
both spatial diffusion and momentum losses, thus being suitable to
test both the energy-loss and the re-acceleration steady-state scheme,
respectively. We first apply the test to the energy-loss scheme. The
only term that is neglected in
\mbox{Eq. (\ref{EqPropagationSteadyEnLoss})} is thus the catastrophic
loss term, which represents a simple ODE in itself.

To allow for an analytical solution, we assume that the spatial and
the momentum dimension decouple from each other. That is, we will
investigate problems with a solution of the form:
\begin{equation}
  \psi = \psi_{\vec{r}}(\vec{r}) \psi_p(p)
\end{equation}
For the spatial diffusion we assume the same spatially constant
coefficients as for the first test in the previous section. For this
case one can find the solution
\begin{align}
  \psi_{\vec{r}}
  &=
  \cos\left(\frac{\pi}{2}\frac{x}{R}\right)
  \cos\left(\frac{\pi}{2}\frac{y}{R}\right)
  \cos\left(\frac{\pi}{2}\frac{z}{H}\right)
  \nonumber\\
  \psi_p
  &=
  C p^{-b} + \frac{s_0}{a} \frac{1}{n-b-1+s} p^{1-s-n}
  \qquad\textnormal{with}\qquad
  b = \frac{a n - \lambda^2_0}{a}
  \label{EqTestFullSolution}
\end{align}
For a specific derivation of this result see
\ref{SecAppendAnalyt}. There also the specific simplifications that
are made are discussed in detail. Some of the transport parameters
need to be fixed to simplify the solution; others are still free.

In particular we are using a diffusion tensor, with entries that
depend on momentum $p$ as: $D_{xx} = D_{yy} = 10^{24} p^{\delta}$
m$^2$\,s$^{-1}$ and $D_{zz} = 10^{26} p^{\delta}$ m$^2$\,s$^{-1}$,
where $\delta = 0.33$. This is actually in rough agreement with
typical values used for Galactic Cosmic Ray propagation. At the same
time we use a power-law for the loss term $\dot p$ and for the source
term $s(p)$ as already discussed in Sec. \ref{SecTestMom}. In the
present case, however, we use the parameters: $a = 15 \lambda_0^2 /8$,
$n=1+\delta$, $s_0 = 10^{-10}$ and $s=1.2$. Some of these choices help
finding an analytical solution, whereas alternative parameters
constitute another independent test.

For the spatial extent of the numerical domain we choose $R=20$ kpc
and $Z=3$ kpc leading to $\lambda^2_0 \simeq 0.324$ pc$^{-2}$ or
$\lambda^2_0 \simeq 3.398\cdot10^{-34}$ m$^{-2}$ (for the specific
form of $\lambda_0$ see \ref{SecAppendAnalyt}).

This test was run with different resolutions in normal and
momentum-space. For this we, again, found second order behaviour as
expected for the code. In particular we found for 33 gridpoints in
each of the spatial dimensions and approximately 100 gridpoints in
momentum space that the deviation from the analytical solution
(computed via the relative l2 error) was on the $0.1\%$ level for
the steady state solver. This finally shows that also the full
transport equation is solved efficiently and highly accurately.

\begin{figure}
  \begin{center}
    \setlength{\unitlength}{0.008cm}
    \begin{picture}(1100,862)(-100,-100)
      \put(390,-70){$p$ [MeV/$c$]}%
      \put(-70,380){\rotatebox{90}
        {$\psi$}}%
      \includegraphics[width=1000\unitlength]{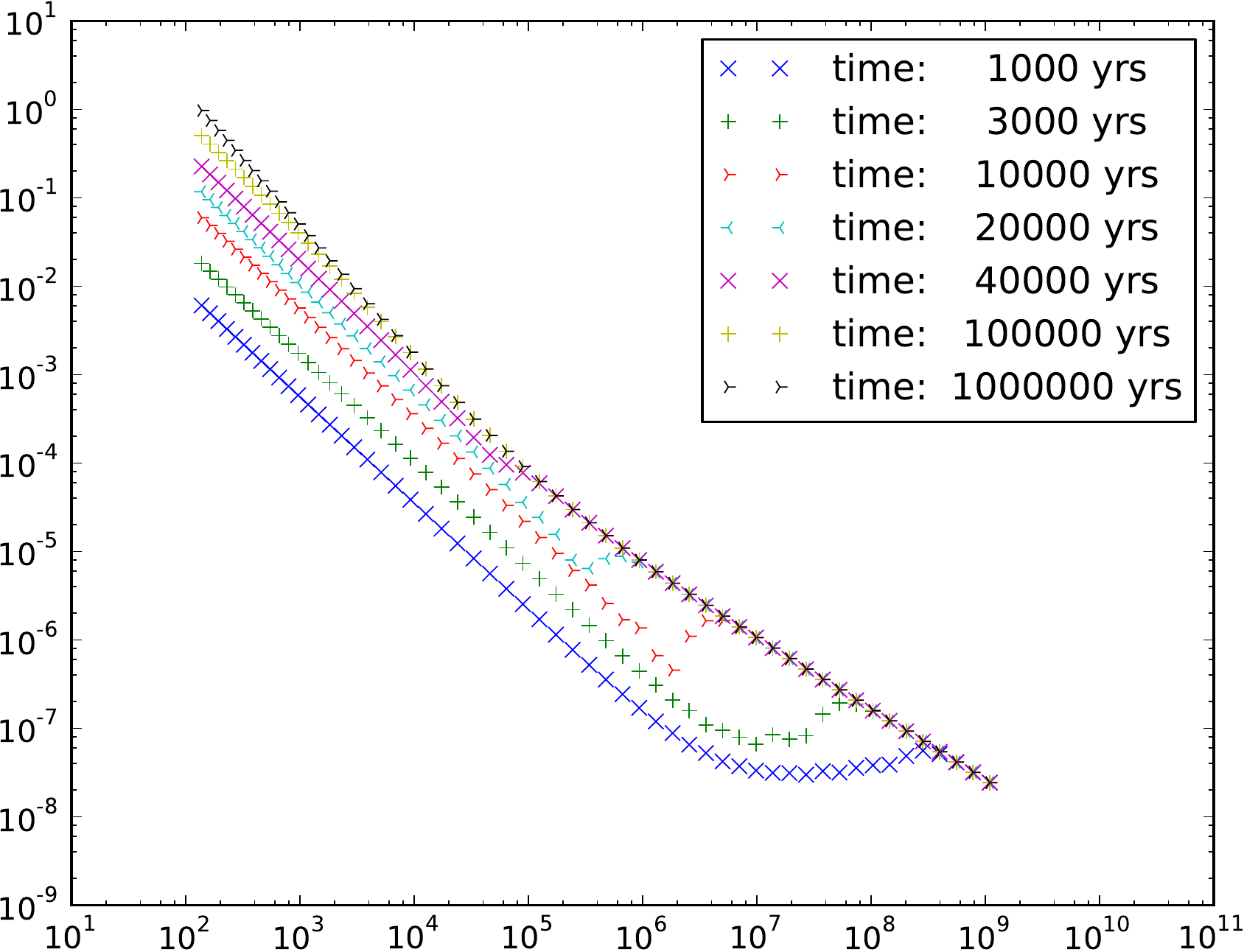}
    \end{picture}
  \end{center}
  \caption{\label{FigTestCodeTimeConv} Results for the overall test of
    the code using the time-dependant version. Results are shown for
    different times, where we show the converged solution for a time
    of $10^6$ years.}
\end{figure}

We also tested the same configuration using the time-dependant
solver. In that case we initialised the Cosmic Ray distribution function with
the analytical value at the highest momentum throughout the whole
numerical domain. This value was used as a boundary condition for the
whole simulation run. From this we also found convergence to the
analytical solution. However, we also found that convergence can take
a very long time. This is depicted in
Fig. \ref{FigTestCodeTimeConv}. As expected from the fact that we only
consider momentum losses it is obvious that the correct solution is
advanced from high to low momenta. Also it is obvious that convergence
is taking longer for lower momenta which is due to the fact that the
energy loss rate decreases with decreasing momentum.

From Fig. \ref{FigTestCodeTimeConv} one can also see some fluctuations
below the momentum, for which the steady state solution has already
been reached. This is due to the steep slope in this region. Such
fluctuations can be avoided by using a slope limiter that would
suppress such fluctuations. We also tested a different
time-integration scheme using operator splitting. In that case the
momentum losses were handled using a variation of the semi-Lagrangian
scheme presented in \citet{ZerroukatEtAl2006IJNMF51_1097}. For this,
such fluctuations did indeed not occur. The resulting solver,
however, was not as accurate as the one presented here anymore. Also
one has to bear in mind that such jumps are not to be expected for a
physical application to Galactic Cosmic Ray transport, where we are
mostly dealing with smooth power-laws.

In this case the solution is shown as a function of momentum instead
of energy because the analytical solution is also a function of
momentum. In particular the break between the two power-laws is
clearly visible in the region around $p=100$ GeV/$c$ which is
determine from the specific choice of $C$ in
Eq. (\ref{EqTestFullSolution}). Time is given in units of years, where
a single timestep is 100 years. Note, however, that this constitutes
only an exemplary test without deep physical meaning. The important
point, however, is the effective number of timesteps, which can
readily be deduced from Fig. \ref{FigTestCodeTimeConv} to be in excess
of 1000 steps. From a runtime evaluation of the l2-error we find in
particular that a steady state solution is reached at about $t=500000$
years or 5000 timesteps.

Finally, we simulated the same model with the re-acceleration
scheme. Even though this scheme has been implemented for the solution
of Cosmic Ray transport problems with re-acceleration, it is also
applicable to a problem with just energy losses. In this case we
solved the model problem for $N$=17, 33, 65 gridpoints in each of the
four dimensions. While the solver turned out to be indeed suitable for
solving this model problem we observed at the same time that the
momentum part of the scheme is in contrast to the energy-loss scheme
just of first order. This is due to the discretisation of the momentum
loss term in Eq. \ref{EqMomUpwind} being of first order -- as is
common in Galactic Cosmic Ray propagation codes like, e.g.,
\textsc{Galprop}. This results in an error of a several percent in
comparison to the analytical solution at $N=65$ in contrast to an
error of a fraction of a percent for the energy-loss solver.

While this might seem problematic at first sight one has to bear three
important points in mind. First, an error of a few percent in energy
is a very decent result for Cosmic Ray transport models. Secondly, in
the present test the loss-rate is of critical importance for the
solution, while -- with the exception of electrons -- the energy
losses only have a significant influence at the low energy end of the
Cosmic Ray spectrum. This is also reflected in the comparison test
discussed later. And third, the Cosmic Ray distribution function is
very smooth in momentum space, thus, not requiring excessively high
resolution there.

From these tests we can draw several important conclusions: The code
works as expected for the analytical tests presented here. This is
true for both the steady state version and the time-integration
version of the code, which yield identical results when the steady
state is reached for the time-integrating solver.

When it comes to the decision which solver to use the answer is
therefore obvious: Whenever interested in a steady state solution the
steady state solver is the preferred one. This is due to the fact that
it is significantly faster than a time-integrating solver that also
has to check for convergence to a steady state result. For the overall
test of the code the time-integrating scheme obviously needs more than
$10^3$ time steps to converge (at $t=10^5$ years the results are not
converged yet as can be seen in Fig. \ref{FigTestCodeTimeConv}). One
also has to keep in mind that for higher resolutions also smaller
timesteps are needed, thus, increasing integration time even
further. Taking into account that the numerical cost of finding the
steady state solution is of the same order as the cost of computing a
single timestep in the time-integrating version of the solver the
steady state solver obviously performs significantly faster in this
case.

For time variable problems (as, e.g, when using variable sources) the
only choice is to use the time-integrating solver, where the steady
state solver might be used to find an appropriate initial
condition. With this remark we can now discuss a first application of
the new code to the Galactic Cosmic Ray transport problem.

\section{Galactic Cosmic Ray propagation}
\label{SecExample}
In this section we will discuss three particular example
scenarios. First, we will investigate a case that has also been used
by \citet{WernerEtAl2013arXiv1308_2829W} to test the \textsc{Galprop}
solver with high resolution simulations in three spatial
dimensions. Then we will verify the re-acceleration solver by a
comparison to \textsc{Galprop} simulations for a physically motivated
model including re-acceleration and to an energy-loss problem solved
with the energy-loss scheme. Finally, we will investigate the Galactic
Cosmic Ray propagation problem for electrons and protons considering
the effect of a time variable source.

\subsection{Comparison to \textsc{Galprop}}
In the first model we investigate an azimuthally symmetric problem for
Galactic Cosmic Ray proton propagation as it was introduced in
\citet{WernerEtAl2013arXiv1308_2829W}. In this model we use a
simulation domain with the extent $x,y = -15\dots15$ kpc and
$z=-4\dots4$ kpc. In momentum space we use a logarithmic grid for
kinetic energy in the range $E_{kin} = 100\dots10^9$\,MeV, which is
discretised using the same number of gridpoints as for each spatial
dimension. In this case the upper boundary in momentum space is
significantly higher than in \citet{WernerEtAl2013arXiv1308_2829W}
where the current choice is motivated by the boundary conditions at
the highest energies. Due to the fact that we are dealing exclusively
with energy loss processes in momentum space we have an inflow
boundary there. Such a boundary can easily lead to numerical
instabilities. Therefore we use the assumption that the Cosmic Ray
density above $10^9$\,MeV is negligible, which yields the numerically
stable boundary condition $\psi(p=p_N) = 0$.

In these simulations we focus on the propagation of Galactic Cosmic
Ray protons, where simulations with different resolutions were run to
check for convergence of the results. Results are compared to
computations using \textsc{Galprop} with the same parameters. Here we
compare simulations with $N=33, 65, 129$ where $N$ is the number of
gridpoints in each dimension including the momentum dimension. For
\textsc{Galprop} we did non run the simulations with $N=129$ due to
excessive memory demand and long computation time. For $N=65$ we found
the steady state solver presented here to be more than a factor of ten
faster than the equivalent simulation with the \textsc{Galprop}
solver.

\begin{figure}
  \begin{center}
    \setlength{\unitlength}{0.008cm}
    \begin{picture}(1100,866)(-100,-100)
      \put(420,-70){$E$ [MeV]}%
      \put(-70,320){\rotatebox{90}
        {$E_{kin}^2 \psi$}}%
      \includegraphics[width=1000\unitlength]{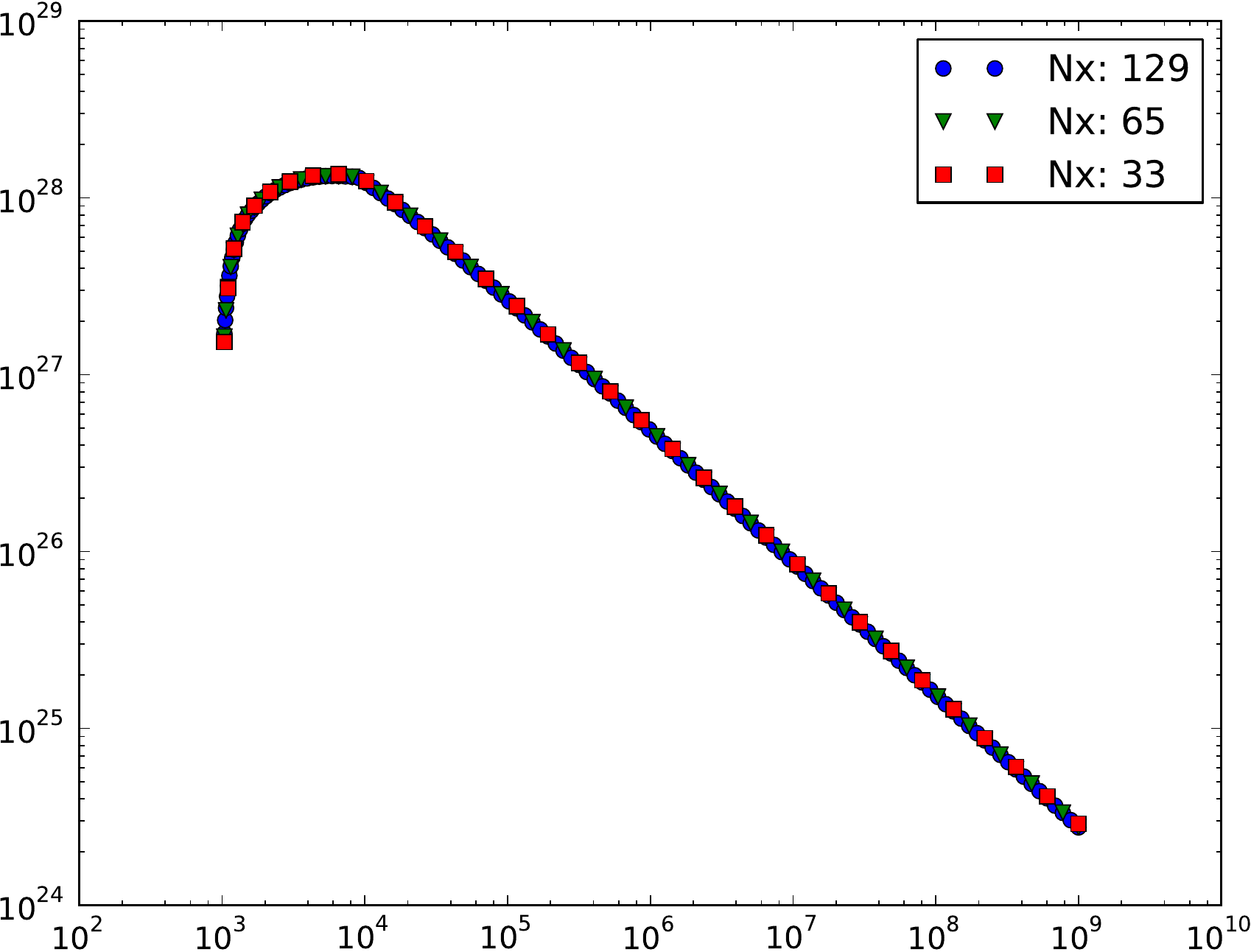}
    \end{picture}
  \end{center}
  \caption{\label{FigConvNew}Proton spectrum at the Galactic
    centre for different resolutions as indicated in the
    figure. Solutions were computed with the steady state
    solver. Results are given in arbitrary units.}
\end{figure}

Convergence of the results was studied by a comparison to the
simulation with $N=129$. We found that the quality of the result for
this simulation is similar for the new solver and the \textsc{Galprop}
solver. The different runs computed with the steady state solver are
compared in Fig. \ref{FigConvNew}, where the flux at the position of
the Galactic centre is shown as computed by the solver. Results are
given in non-normalised form. To compute values that can be compared
with experimental data results are usually normalised. For this the
numerical data are rescaled so that the flux is identical to the
observed one at the position of Earth at a specific energy (typically
around 100~GeV). When the position of Earth does not coincide with a
numerical gridpoint an interpolation procedure is used for the
normalisation. This, however, might have an influence on the estimate
of the quality of the convergence, because the position of Earth
relative to the nearest gridpoint might change with changing spatial
resolution.

The deviations are on the order of 10\% for a resolution of $N=33$ and
of the order of a few percent for $N=65$ for the new solver. When
comparing the normalised values the deviation turns out to be about
twice as large. For the \textsc{Galprop} run we only compared the
$N=33$ to the $N=65$ which show about the same deviation as for the
new solver. A comparison between the two solvers also shows good
agreement, within a few percent deviation. This reflect the fact that
the deviations are just due to the discretisation error of the
methods, which is similar in both cases.

This shows that the results obtained using the solver presented here
are indeed sufficiently accurate to be applied to actual physical
problems. The new quality of this solver can be described two-fold: it
does not need any tests of the convergence criteria to be studied,
while being more than a factor of 10 faster than the \textsc{Galprop}
solver. Finally the present solver has also been written in a fully
MPI parallel way, thus, making use also of distributed memory
machines, advantageous for very high resolution simulations.

\subsection{Re-acceleration runs}
Yet another verification of the re-acceleration solver is investigated
using two different simulations, for which we compare results by the
re-acceleration solver to those from the energy-loss scheme and those
computed using the \textsc{Galprop} code, respectively. We considered
two different standard \textsc{Galprop} simulation setups described in
\citet{StrongEtAl2010ApJ722L_58S}.

As a first test we chose model z04LMPDS, which is a setup without
re-acceleration, where only protons and electrons are considered. In
this case we compared simulations with both the energy-loss and the
re-acceleration steady state scheme. Both simulations are in fairly
good agreement with a relative l2-deviation of the proton spectrum at
Earth of 16\% at $N=33$ and 8\% at $N=65$ with $N$ the number of
gridpoints in each dimension. In this case we only observe deviations
above the percent level below energies of about 2\,GeV, which is
likely due to the fact that energy losses become increasingly
important for low energy  protons only. This localisation of the
deviation becomes also apparent, when computing the deviation using
the l1-norm $\Delta_{l1}$ instead:
\begin{equation}
  \Delta_{l1}^N = \frac{1}{N}
  \sum_{l=0}^{N-1}
  \frac{\left|\psi_2(p_l) - \psi_1(p_l)\right|}{\left|\psi_1(p_l)\right|}
\end{equation}
where we used the solution obtained using the energy-loss scheme as
reference solution $\psi_1$. In this case we find that
$\Delta^{33}_{l1} = 1.7$\% for $N=33$ and $\Delta_{l1}^{65} = 0.9$\%
for $N=65$. Taking into account that the l2-deviation puts more weight
on localised errors especially at low energies it becomes apparent
that we have a very good agreement.

For electrons we found somewhat larger deviations, as expected from
the larger impact of energy losses. In this case we determined errors
of $\Delta^{33}_{l1} = 28.9$\%, $\Delta^{65}_{l1} = 16.9$\%,
$\Delta^{33}_{l2} = 11.8$\% and $\Delta^{65}_{l2} = 7.2$\%. These
errors reflect the observation also drawn from a visual inspection of
the spectra that we see some deviation at all energies with larger
differences towards high energies. This indeed supports the idea that
the results deviate most strongly, where energy losses become
relevant. Taking into account that the discretisation in both schemes
are quite different the agreement is fairly good. In particular
the decrease of the error for higher resolution shows that both
schemes converge towards the same solution. Additionally, we found for
this test that the energy-loss solver is about a factor of 3 faster
than the re-acceleration solver.

As the second example we used model z04LMS, taking only protons into
account, to compare the results of a simulation with re-acceleration
to the corresponding results found using the \textsc{Galprop} code. In
this case we find for $N=33$ a relative l2 deviation of the proton
spectrum at Earth of about 5,8\%, while at $N=65$ we find a deviation
of about 1.9\%. In this case the l1 deviation is with
$\Delta^{33}_{l1} = 4.1$\% and $\Delta^{33}_{l1} = 0.95$\% very
similar to the l2 deviation showing that the error is not localised to
a specific energy range in this case. This was to be expected due to
the fact that the discretisation of the energy loss terms is of first
order in both cases.  This shows that the \textsc{Picard} code yields
reliable results very efficiently also in the presence of
re-acceleration. In the final test we apply the energy loss scheme to
a dynamical problem involving electrons and protons.

\subsection{Dynamical effects}
In the final test we apply the new energy-loss solver to a
specific Galactic propagation problem. For this we used the same
parameters as in \citet{StrongEtAl2010ApJ722L_58S}, where the
simulation domain extents from -20 to 20 kpc in the $x-y$ plane and
from -4 to 4 perpendicular to the Galactic plane. All other parameters
can be found in table 1 of
\citet{StrongEtAl2010ApJ722L_58S}\footnote{Designated as \emph{model
    z04LMPDS}.}.

\begin{figure*}
  \begin{center}
    \setlength{\unitlength}{0.00041\textwidth}
    \begin{picture}(1143,1100)(-100,-100)
      \put(470,-70){\small$x$ [kpc]}%
      \put(-70,420){\rotatebox{90}
        {$y$ [kpc]}}%
      \includegraphics[height=1000\unitlength]{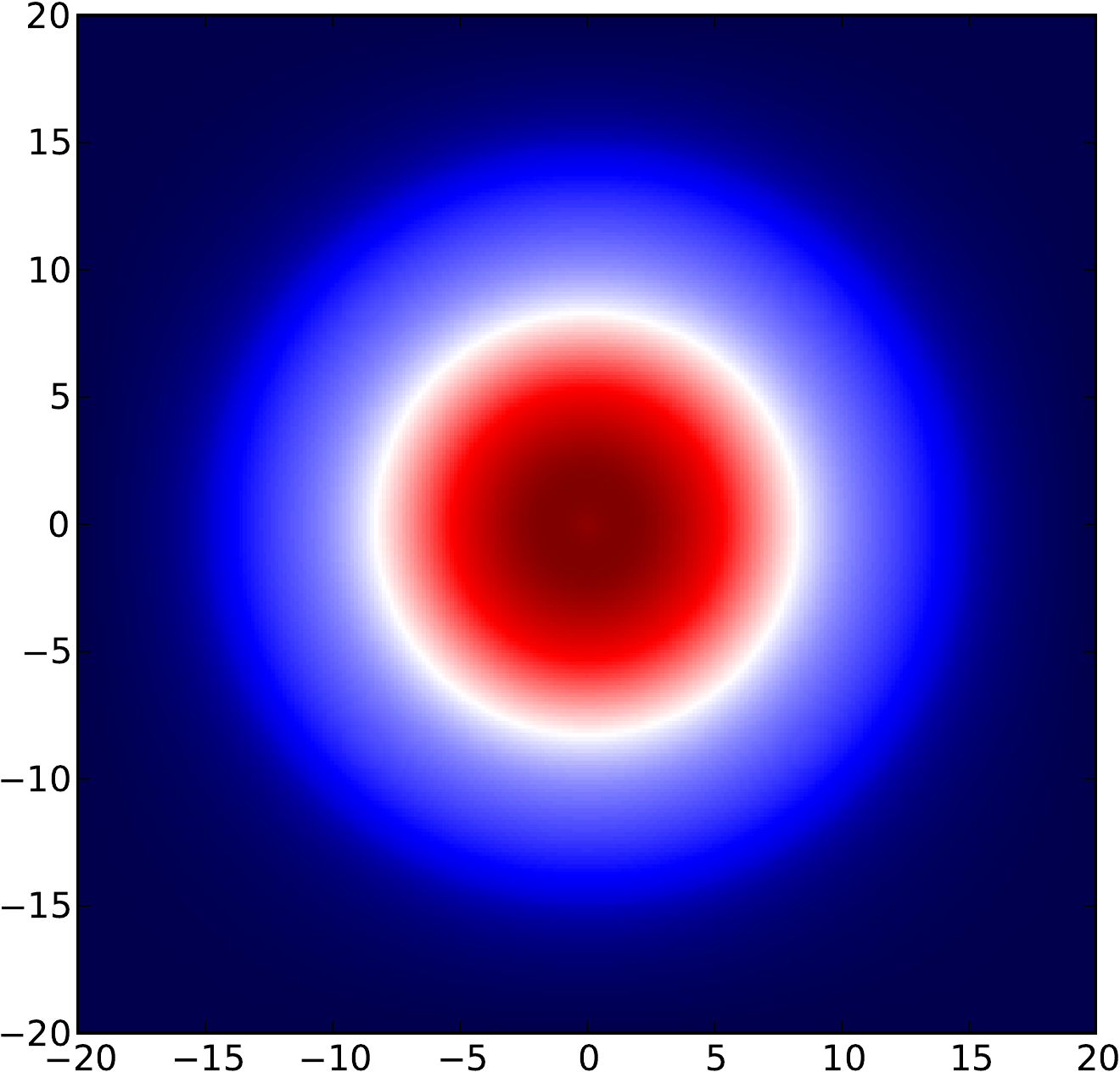}
    \end{picture}
    ~\hfill
    \begin{picture}(1217,1118)(0,-100)
      \put(470,-70){$x$ [kpc]}%
      \includegraphics[height=1018\unitlength]{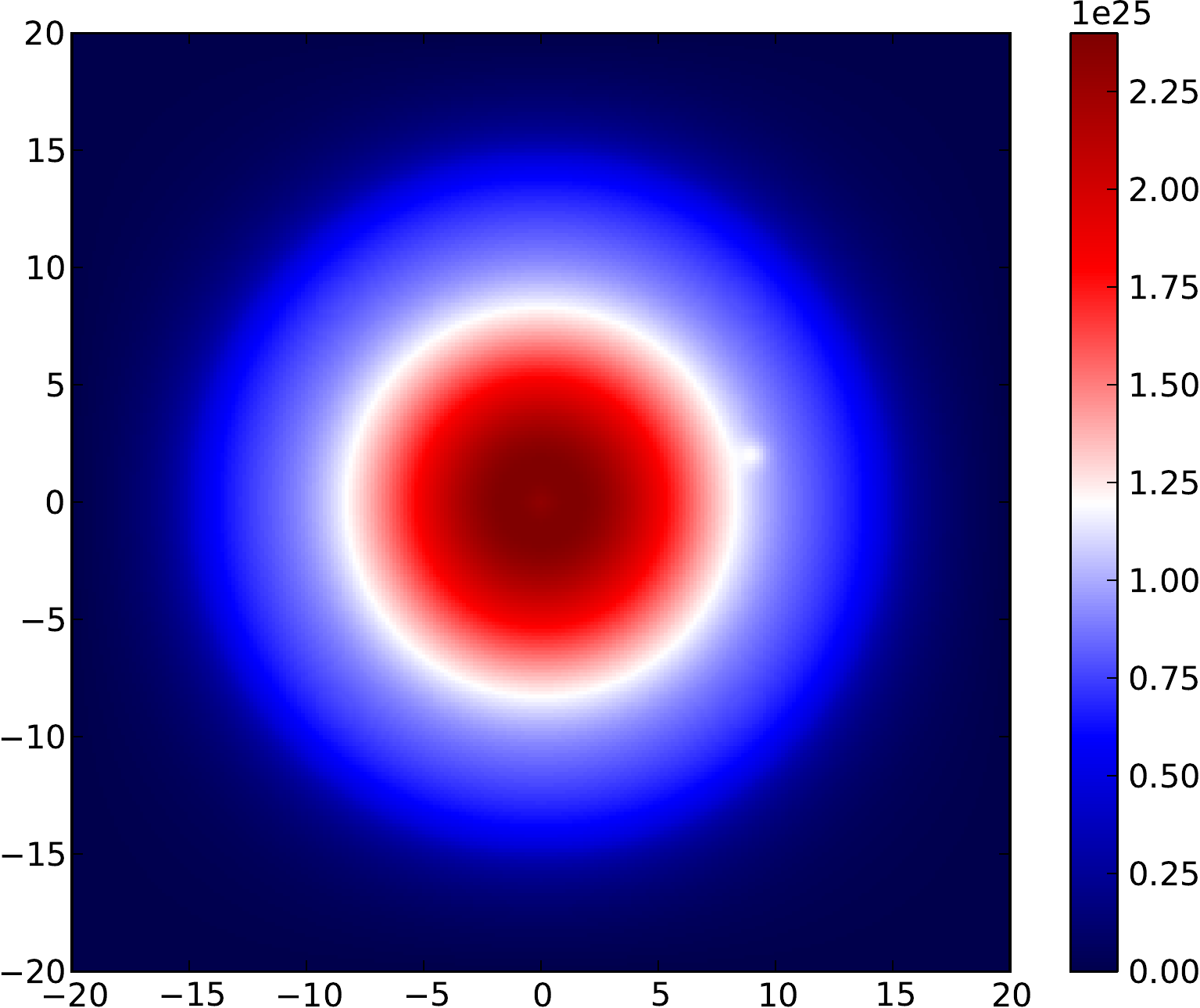}
    \end{picture}
  \end{center}
  \caption{\label{FigProtonDist}Proton flux in the Galactic plane for
    an energy of $\sim$ 103 TeV. All cells are given with a constant
    colour. On the left results are shown without the additional
    source and on the right the source was active for about $10^4$
    years. The flux is given as $F = E_{kin}^2 \psi$ in arbitrary
    units.}
\end{figure*}

As a difference we presently only consider electrons and protons. At
the same time we did the simulation in three spatial dimensions
instead of using an azimuthally symmetric setup. This was done for two
reasons. First we want to show the capability of the new scheme to
recover the azimuthal symmetry using a Cartesian 3D spatial
grid. Secondly we included a dynamical source in the vicinity of Earth
at coordinates $(x,y,z) = (9,2,0)$ in units of kpc (where Earth is
located at $(x,y,z) = (8.5,0,0)$ kpc). This source is switched on 100
years after the start of the simulation and is active for $10^4$
years. By this we of course brake the azimuthal symmetry of the setup,
thus, forbidding a spatial 2D simulation.

For this simulation we first computed the steady state solution and
then switched on the time integration scheme using the steady state
solution as initial condition.  In Fig. \ref{FigProtonDist} we show
the initial spatial distribution of \mbox{$\sim$103 TeV} Galactic
Cosmic Ray protons in the Galactic plane on the left hand side. In
this Fig.  all cells are shown individually, each with constant
colour. The high resolution is apparent from the Fig., where we used a
cell size of only 0.156 kpc in the $x$-$y$-plane. Also the results
show indeed the azimuthal symmetry expected from the azimuthally
symmetric source distribution.

\begin{figure*}
  \begin{center}
    \setlength{\unitlength}{0.00041\textwidth}
    \begin{picture}(1143,1100)(-100,-100)
      \put(470,-70){\small$x$ [kpc]}%
      \put(-70,420){\rotatebox{90}
        {$y$ [kpc]}}%
      \includegraphics[height=1000\unitlength]{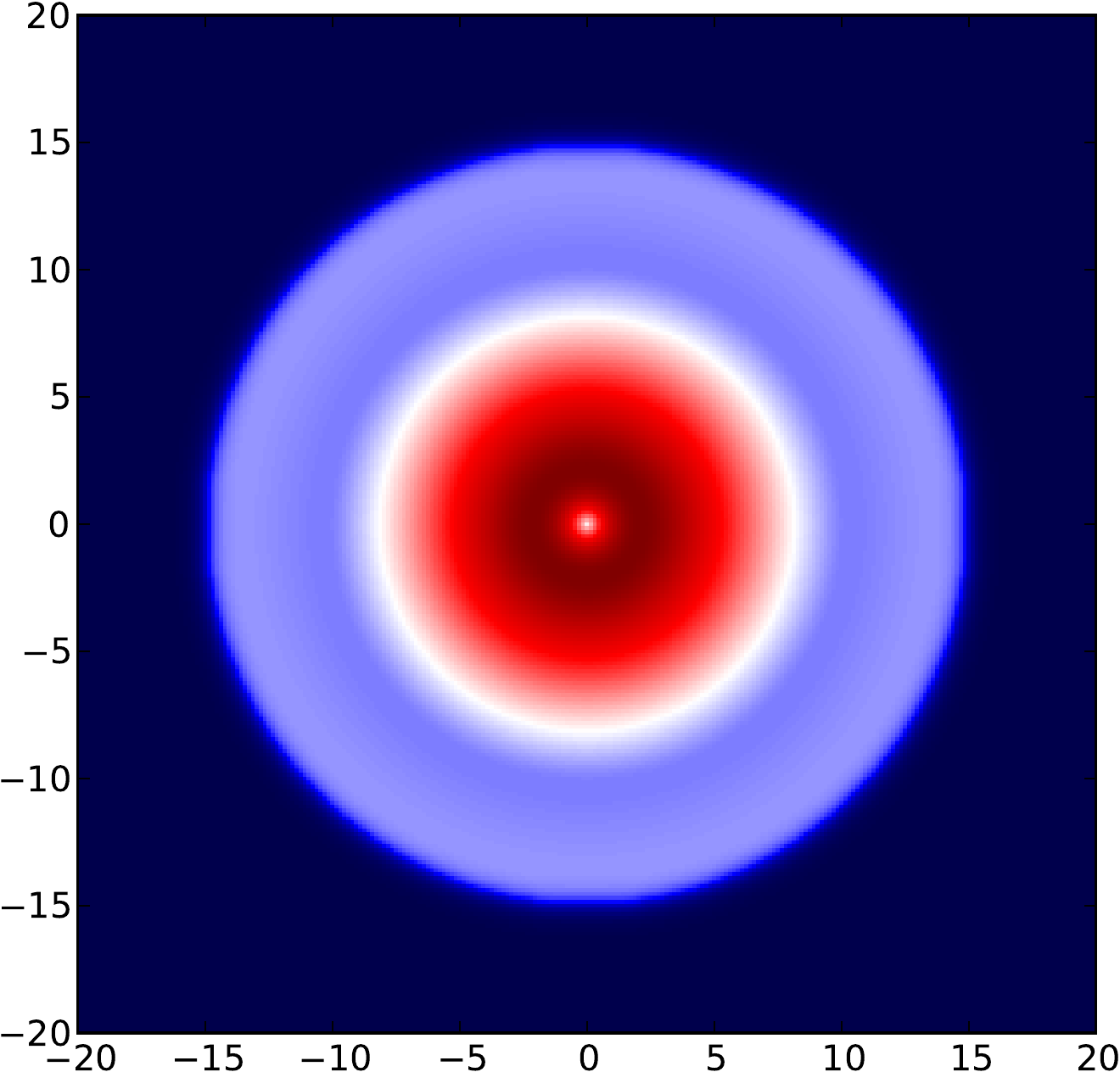}
    \end{picture}
    ~\hfill
    \begin{picture}(1217,1118)(0,-100)
      \put(470,-70){$x$ [kpc]}%
      \includegraphics[height=1018\unitlength]{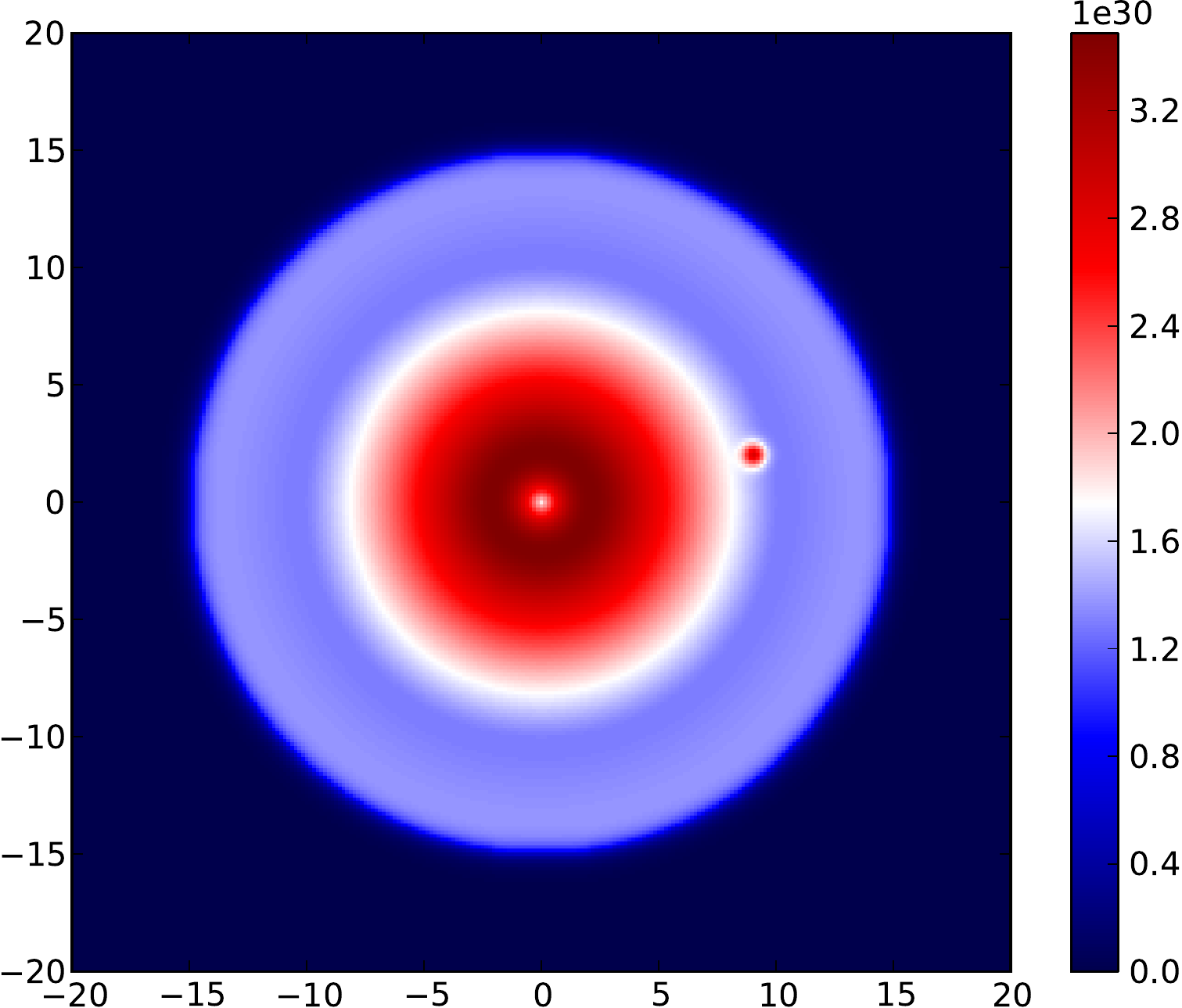}
    \end{picture}
  \end{center}
  \caption{\label{FigElecDist} Same as Fig. \ref{FigProtonDist} only
    shown for electrons with an energy of $\sim$5.05 TeV.}
\end{figure*}

In the same Fig. we also show the result in the presence of the
additional source, which is clearly visible as the localised
peak. These results are shown at a time of $t=10^4$ years near the end
of the active phase of the additional source. For protons, however,
the effect of the localised source is rather weak resulting in a local
increase of the Cosmic Ray density at $E\simeq103$~TeV of about
13.5\%. For electrons this effect is much more pronounced and is still
relevant at significantly lower energies as expected \citep[see,
  e.g.][]{PohlEsposito1998ApJ507_327}.

In Fig. \ref{FigElecDist} we show the electron distribution at
$E\simeq5.05$~TeV at the same times as used in
Fig. \ref{FigProtonDist}. Obviously the source has a much higher
impact for electrons at an even lower energy than in the proton
case. In particular we find a local increase of $\sim$90\% for the
electron flux here, which becomes even higher for higher energies.
\begin{figure*}
  \begin{center}
    \setlength{\unitlength}{0.00036\textwidth}
    \begin{picture}(1172,1100)(-100,0)
      \put(-70,420){\rotatebox{90}
        {$y$ [kpc]}}%
      \includegraphics[height=1000\unitlength]{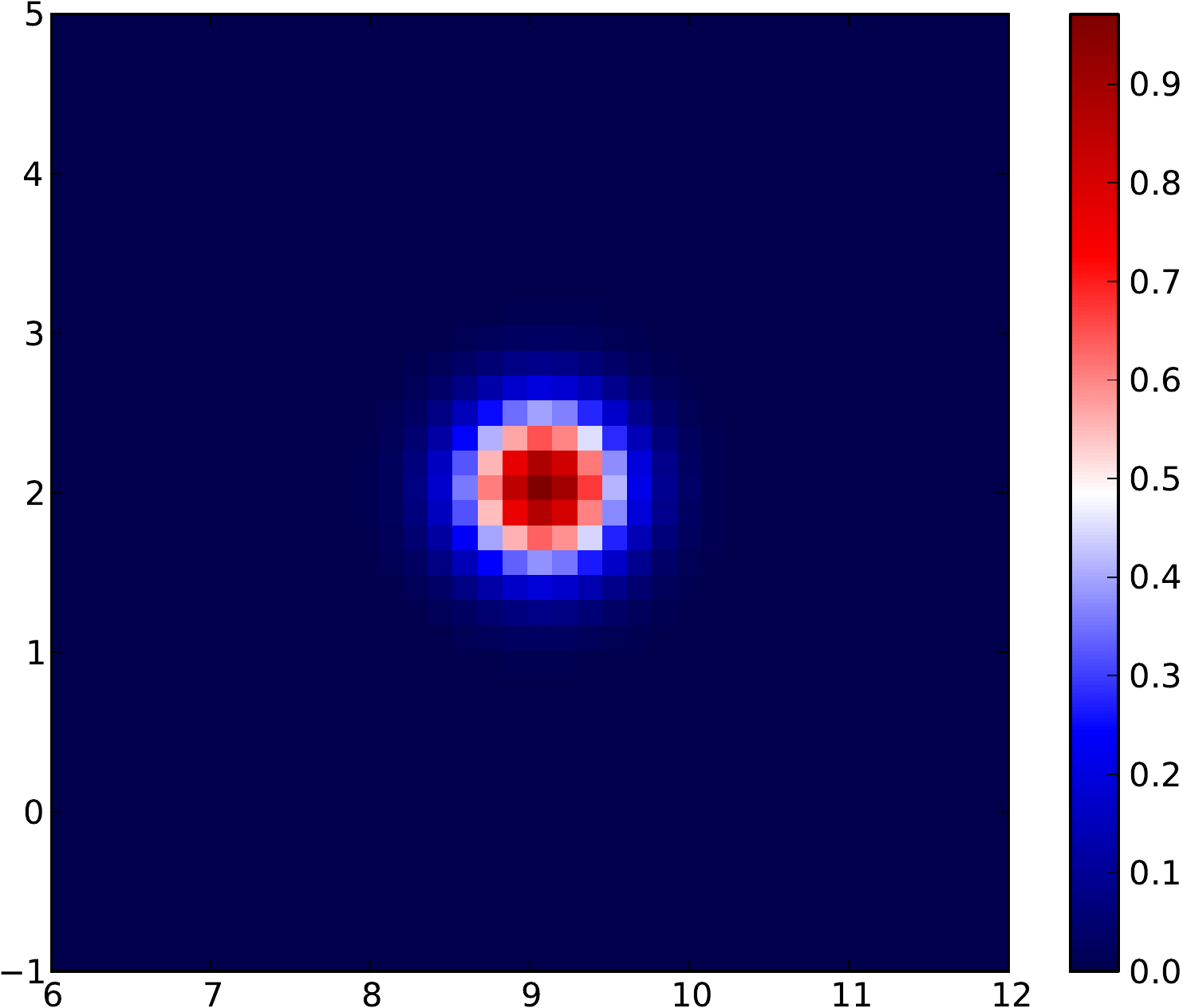}
    \end{picture}
    ~\hfill
    \begin{picture}(1192,1000)(0,0)
      \includegraphics[height=1000\unitlength]{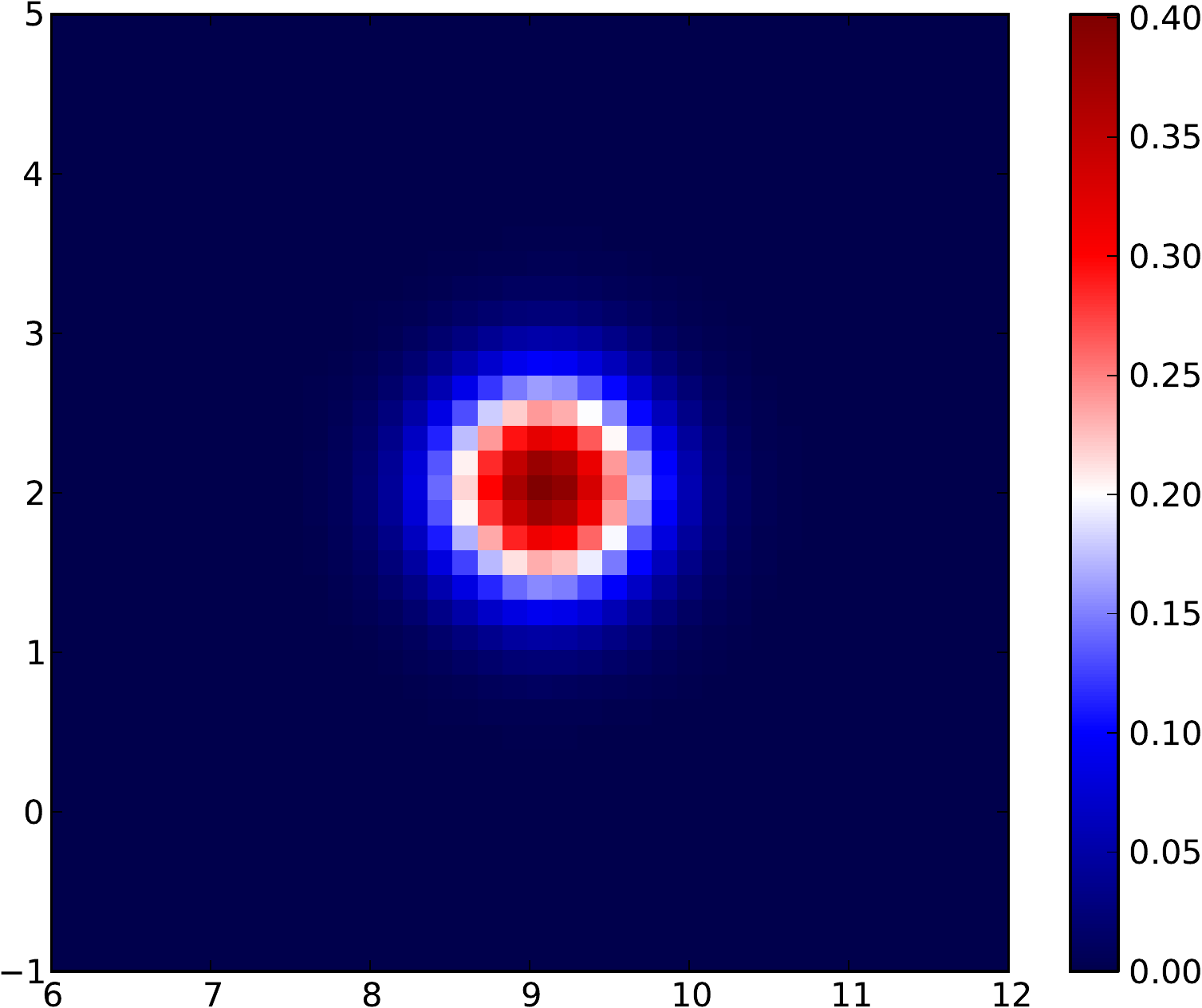}
    \end{picture}
    ~\\~\\
    \begin{picture}(1313,1100)(-100,-100)
      \put(460,-70){\small$x$ [kpc]}%
      \put(-70,420){\rotatebox{90}
        {$y$ [kpc]}}%
      \includegraphics[height=1000\unitlength]{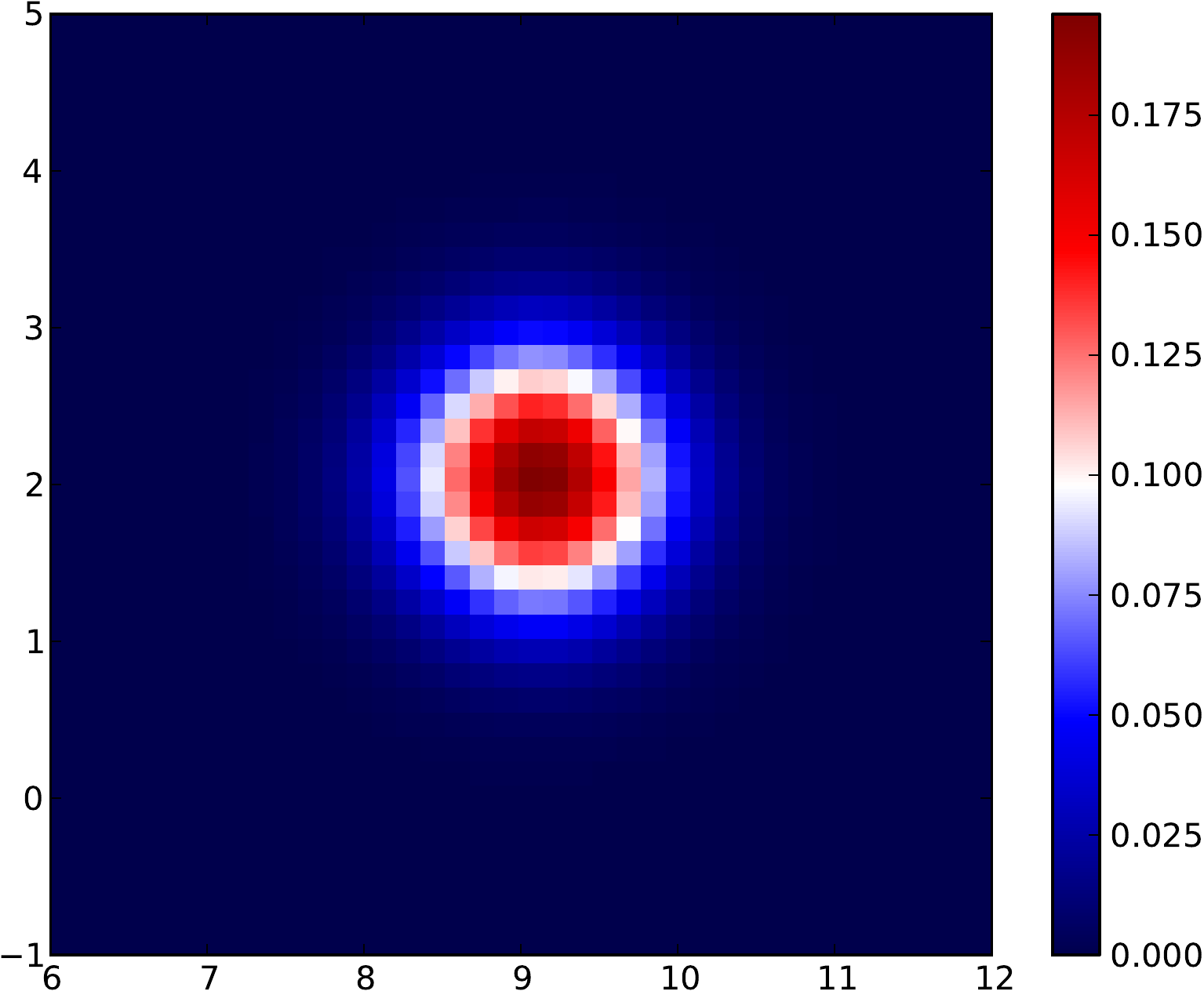}
    \end{picture}
    ~\hfill
    \begin{picture}(1213,1100)(0,-100)
      \put(460,-70){$x$ [kpc]}%
      \includegraphics[height=1000\unitlength]{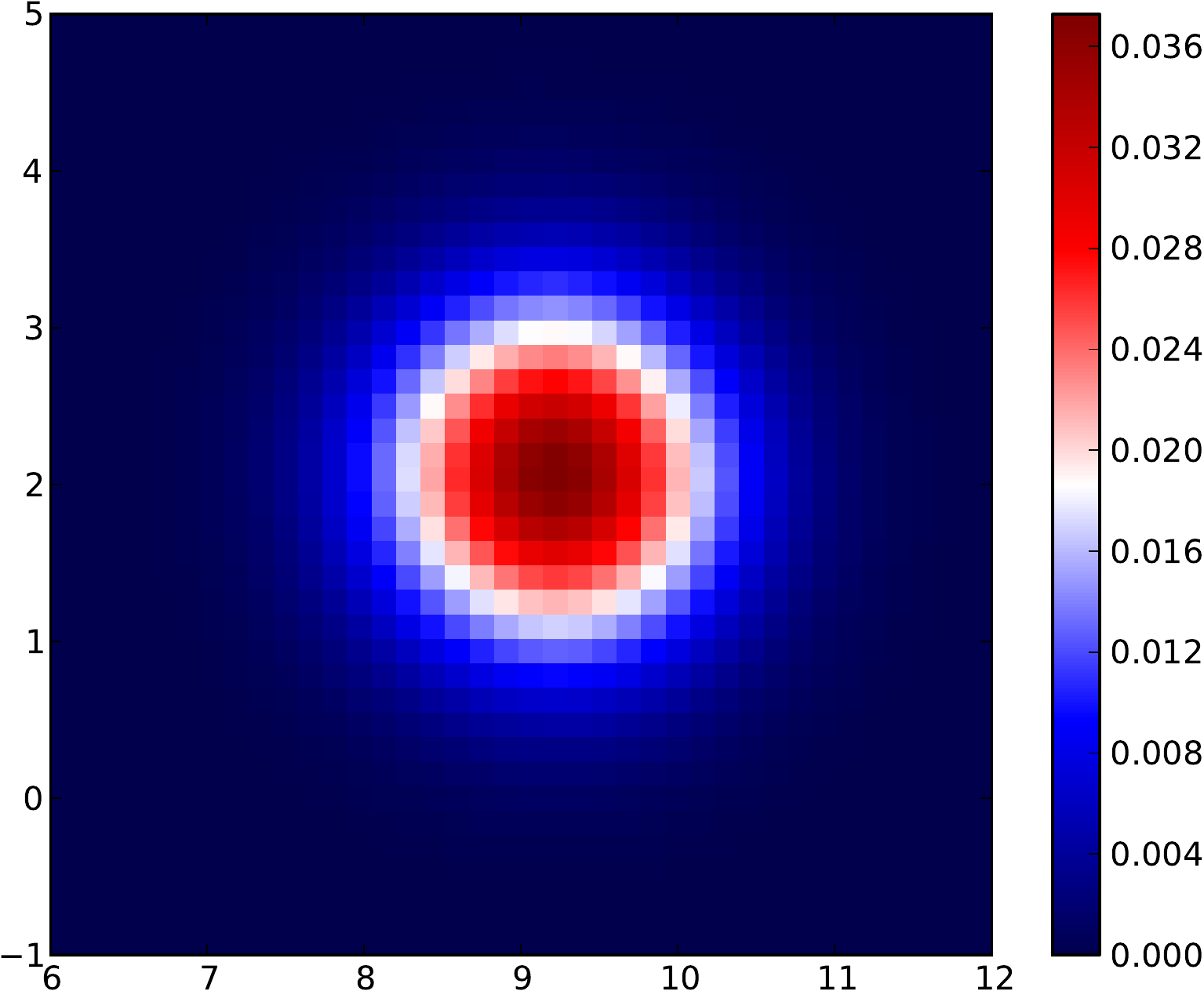}
    \end{picture}
  \end{center}
  \caption{\label{FigElecDistEvol}Residuals for the electron
    distribution at an energy of $\sim$5.05 TeV for different times,
    where $t=10^4, 2\cdot10^4, 3\cdot10^4, 5\cdot10^5$ years from the
    upper left to the lower right.}
\end{figure*}

For electrons one can now investigate the dynamics of the source by
comparing the flux at different times.  This is shown in
Fig. \ref{FigElecDistEvol} where the residual is plotted for different
times. Here the residual is computed from the comparison with the
result at $t=0$, where no additional source was present. Results are
shown at the end of the active phase of the source and at $10^4$,
$2\cdot10^4$ and $4\cdot10^4$ years after the disappearance of the
additional source. What is obvious from these Figs. is that the effect
of the source disappears from a combination of energy losses and
spatial diffusion. The latter in particular is readily visible from
the plots by the ever larger region affected by the source. Note,
however, that at $t=5\cdot10^5$ years the effect could not be detected
anymore, because the residual even at the position of the sources
approaches the 1\% level.

\begin{figure}
  \centering
    \setlength{\unitlength}{0.008cm}
    \begin{picture}(1100,861)(-100,-100)
      \put(380,-70){$E$ [MeV]}%
      \put(-90,240){\rotatebox{90}
               {$E_{kin}^2 \psi$ [arbitrary units]}}%
      \includegraphics[width=1000\unitlength]{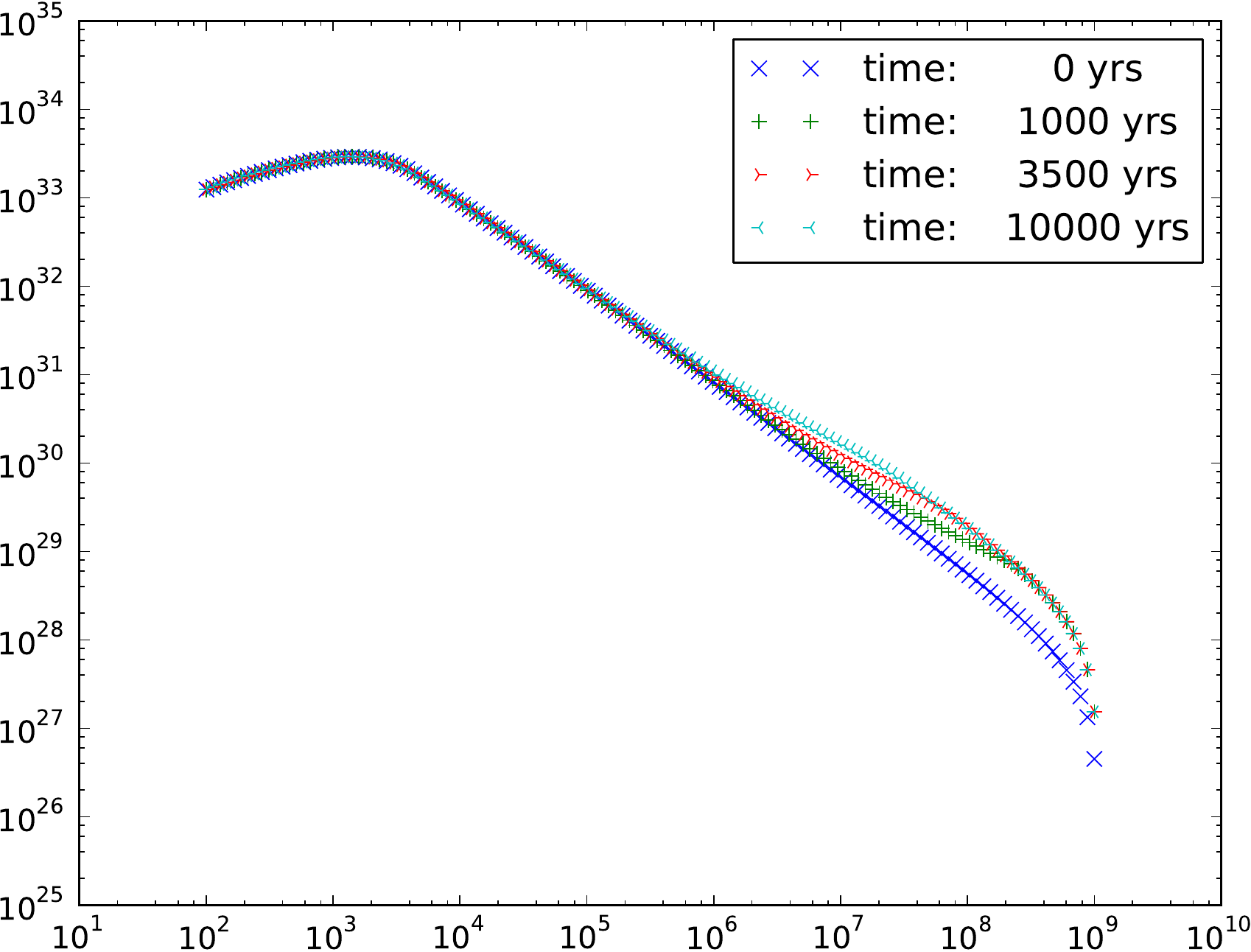}
    \end{picture}
  \caption{\label{FigElecSpecEvol_active}Evolution of the electron
    spectrum for the active phase of the source at the position of the
    source.}
\end{figure}

\begin{figure}
  \centering
    \setlength{\unitlength}{0.008cm}
    \begin{picture}(1100,861)(-100,-100)
      \put(380,-70){$E$ [MeV]}%
      \put(-90,240){\rotatebox{90}
               {$E_{kin}^2 \psi$ [arbitrary units]}}%
      \includegraphics[width=1000\unitlength]{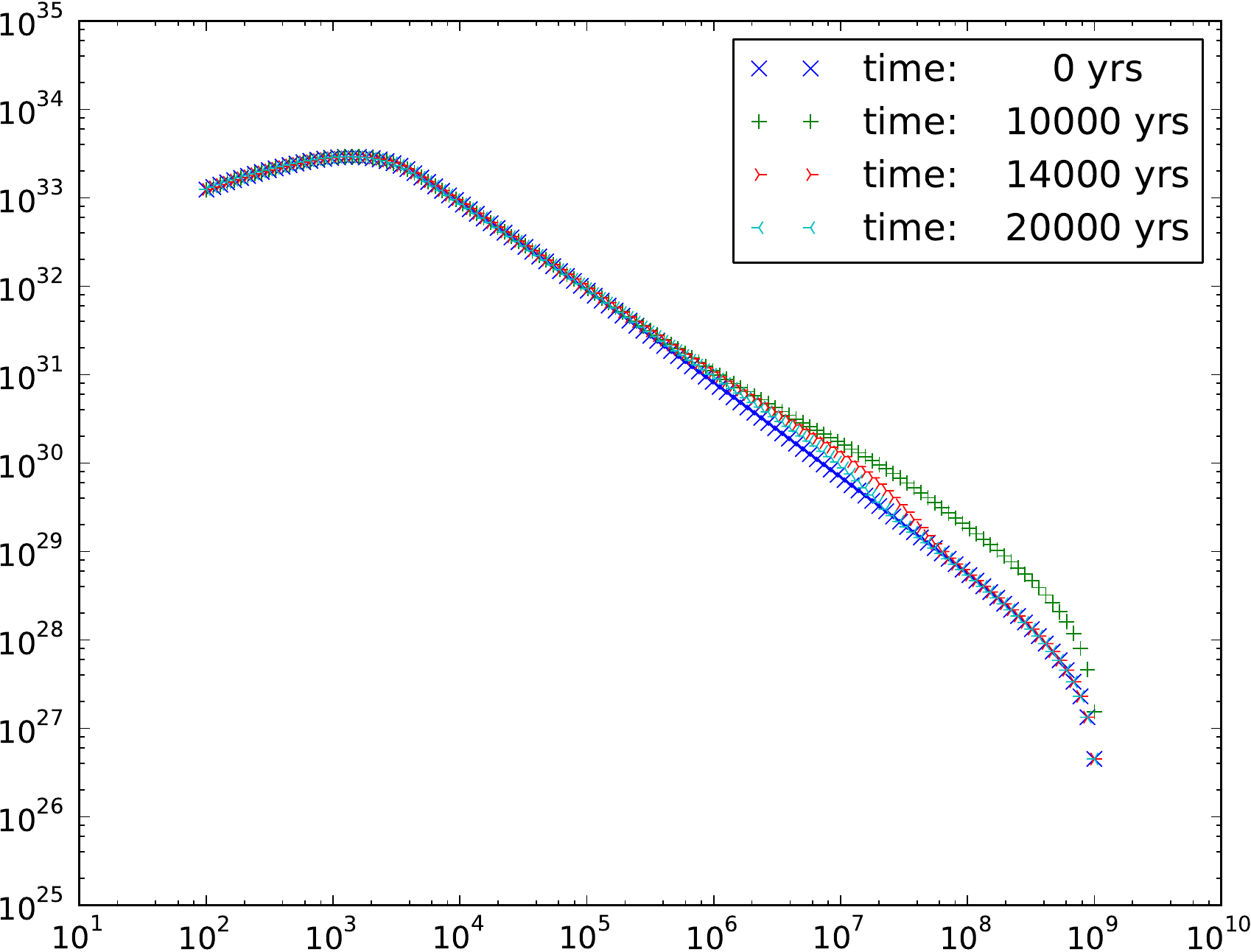}
    \end{picture}
  \caption{\label{FigElecSpecEvol_off}Same as
    Fig. \ref{FigElecSpecEvol_active} also showing the phase after the
    disappearance of the source.}
\end{figure}

This decay can also be visualised by looking at the electron spectrum
at the position of the source shown in
Figs. \ref{FigElecSpecEvol_active} and \ref{FigElecSpecEvol_off}. In
Fig. \ref{FigElecSpecEvol_active} we show the evolution of the spectrum
for the active phase of the source. For high energies quickly a new
equilibrium between losses and the additional source strength is
achieved. This can be seen from the re-appearance of the same
spectral power-law as without the source at a higher flux level. This
re-appearance of the power-low continues to lower energies for
increased lifetimes of the source.

In Fig. \ref{FigElecSpecEvol_off} we show the evolution of the
spectrum after the source became inactive. Apparently, the spectrum
decays back to the undisturbed form most quickly for high energies
again. Thus, the imprint of the previous source is longer visible at
lower energies. At energies below approximately 1~TeV the effect of
the source can hardly be observed. In the present case the effect of
the local source is most obvious at energies above $\sim$10~TeV.  We
re-ran the simulation for 5-times higher source strength, however, for
which we found the spectrum to be affected at lower energies.

\section*{Conclusion}
Here we introduced a new code for the numerical solution for the
Galactic Cosmic Ray propagation problem. The code was presented in
detail in conjunction with tests validating the numerical scheme for
documenting the capabilities of the code. In particular, the
\textsc{Picard} code can compute a steady state solution for the
propagation problem without any adjustments of numerical parameters by
the user. At the same time the code is significantly faster than,
e.g. the \textsc{Galprop} code.

We used the code for an initial physical application, where we
investigated the evolution of a localised Cosmic Ray source on the
proton and electron distributions. We find that, as expected, a
significant impact for electrons and only a minor effect for the
proton distribution. This also shows the applicability of the new
solver to relevant physical problems. In particular the simulation was
run with a spatial resolution of $\sim$0.15 kpc in a fully 3D
setup. This is a genuine quality to the simulation of Galactic Cosmic
Ray propagation, warranted and necessitated by the ever improving
quality of observational data.  Such high resolutions allow for an
investigation of localised sources and 3D structures in the matter
distribution, magnetic fields or radiation fields in our Galaxy.  This
seems to be a very timely development for transferring Cosmic Ray
propagation into a full 3D scenario \citep[see,
  e.g.,][]{GaggeroEtAl2013PhRvL111_021102,JohannessonEtAl2013ICRC,WernerEtAl2013arXiv1308_2829W}.

\section*{Acknowledgments}
I would like to thank O. Reimer, M. Werner, A. Ostermann, P. Csom\'os, K. Reitberger and
K. Egberts for helpful discussions and valuable suggestions. Additionally, A.~W.~Strong helped with useful ideas and important insights into propagation physics.

\begin{appendix}
  \section{Derivation of an analytical testcase}
  \label{SecAppendAnalyt}
  Here we will derive the analytical solution for the test case used
  in Sec. \ref{SecTestOverall}. In particular we are looking for an
  analytical solution to Eq. (\ref{EqPropagationSteadyEnLoss}) where only
  catastrophic losses are neglected. This is of course done to help
  finding an analytical solution, but at the same time this is not a
  severe limitation due to the fact that other terms of the same form
  will still contribute in the end. The PDE we are working with, is thus:
  \begin{equation}
    \label{EqPropagationSteadyAppend}
    -\nabla \cdot \mathcal{D} \nabla \psi
    +
    \frac{\partial \dot p \psi}{\partial p} 
    =
    s(\vec{r}, p, t)
  \end{equation}
  For this there is of course no general analytical solution for three
  spatial dimensions available. So some simplifications will be
  necessary. First we assume that all space and the momentum dimension
  decouple from each other. That is, we will only investigate problems
  with a solution of the form:
  \begin{equation}
    \psi = \psi_x(x) \psi_y(y) \psi_z(z) \psi_p(p) = \psi_r(\vec{r}) \psi_p(p)
  \end{equation}
  Apart from the decoupling of the solution we will also assume that
  the source term $s(\vec{r}, p, t)$ decouples into a spatial and a
  momentum part, where the spatial part is just given by the spatial
  part of the solution:
  \begin{equation}
    s(\vec{r}, p, t) = s_p(p) \psi_r
  \end{equation}
  This is actually the same approach as was used in the first test in
  Sec. \ref{SecTestDiff}.  Just like in that test we will also neglect
  a spatial variation of the components of the the diffusion
  tensor. So these components are only allowed to depend on momentum
  of the particle. In particular we will assume the same momentum
  dependence for all components. That is we can write:
  \begin{equation}
    \mathcal{D} = \mathcal{D}_0 p^{\delta}
  \end{equation}
  where all entries of $\mathcal{D}_0$ are constant. When inserting this
  together with the separation ansatz into the transport equation we
  find:
  \begin{equation}
    -\psi_p  p^{\delta} \nabla \cdot \mathcal{D}_0 \nabla \psi_r
    +
    \psi_r
    \frac{\partial \dot p \psi_p}{\partial p} 
    =
    s_p(p) \psi_r
  \end{equation}
  Dividing this result by $\psi p^{\delta}$ we can decouple the spatial
  part of the PDE and the momentum part of the PDE, where we find
  after some rearrangement:
  \begin{equation}
    \frac{1}{\psi_r}\nabla \cdot \mathcal{D}_0 \nabla \psi_r
    =
    \frac{1}{\psi_p p^{\delta}} \frac{\partial \dot p \psi_p}{\partial p} 
    -
    \frac{1}{\psi_p p^{\delta}} s_p(p) \psi_r = C
  \end{equation}
  Here we assumed that the energy loss rate $\dot p$ does not to
  depend on spatial position. Thus both sides of the equation need to
  be constant because they are identical while depending on different
  variables.
  
  For the spatial diffusion we use the solution used before:
  \begin{equation}
    \psi_r
    =
    \cos\left(\frac{\pi}{2}\frac{x}{R}\right)
    \cos\left(\frac{\pi}{2}\frac{y}{R}\right)
    \cos\left(\frac{\pi}{2}\frac{z}{H}\right)
  \end{equation}
  From this solution we can find the constant $C$ to be:
  \begin{equation}
    C = -\frac{\pi^2}{4}\left(\frac{D_{xx}^0 + D_{yy}^0}{R^2} +
    \frac{D_{zz}^0}{H^2}\right) = -\lambda^2_0
  \end{equation}
  This can now be used to solve the momentum part of the
  equation. That is we have to deal with the differential equation:
  \begin{equation}
    \frac{\partial \dot p \psi_p}{\partial p} 
    =
    -\lambda_0^2 \psi_p p^{\delta} + s_p(p)
  \end{equation}
  For the momentum losses we choose the same general form as in Sec.
  \ref{SecTestMom}. For the energy dependence of the components of the
  diffusion tensor we use $\delta = 1/3$.  Now we first compute the
  solution to the homogeneous problem in momentum space. The
  corresponding ODE is:
  \begin{equation}
    \frac{\partial \dot p \psi_p}{\partial p}
    =
    -\lambda_0^2 p^{\delta} \psi_p
  \end{equation}
  The solution to this is
  \begin{equation}
    \psi_p^h
    =
    p^{-n} 
    \exp
    \left(
    \frac{1}{1+\delta - n} \frac{\lambda_0^2}{a} p^{\delta + 1 - n}
    \right)
  \end{equation}
  where $n$ is the power-law index of the energy loss term.  Now we
  need to address the full equation, i.e., we need to find a
  particular solution. For the momentum source term we, again, use the
  same form as in Sec \ref{SecTestMom}. Some rearrangement then gives:
  \begin{equation}
    \label{EqAppInhom}
    \frac{d \psi_p}{d p}
    =
    \left(\frac{\lambda^2_0}{a} p^{\delta - n} - \frac{n}{p}\right)\psi_p
    -
    \frac{s_0}{a} p^{-(s+n)}
  \end{equation}
  This obviously can not easily be solved, when desiring a somewhat
  general source-terms. In the present case, it is sufficient,
  however, to find a special solution as our analytical test-case. For
  this we use the specific choice $\delta - n = -1$. With this the
  homogeneous solution of Eq. (\ref{EqAppInhom}) computed above
  simplifies significantly and we find it to be a constant power-law
  \begin{equation}
    \psi^h_p =A p^{\frac{\lambda^2_0 - a n}{a}} = A p^{-b}.
  \end{equation}
This, however, can also be integrated to find the inhomogeneous
solution. In this particular case we find for the particular solution:
\begin{equation}
  \psi^p_p = \frac{s_0}{a} \frac{1}{b-s+1-n} p^{1-s-n}
\end{equation}
thus yielding for the overall solution:
\begin{equation}
  \psi_p =C p^{-b} + \frac{s_0}{a} \frac{1}{n-b-1+s} p^{1-s-n}
  \qquad\textnormal{with}\qquad
  b = \frac{a n - \lambda^2_0}{a}
\end{equation}
This solution and the corresponding parameters are still sufficiently
complex to be a good test to the numerical framework. The overall
solution to be compared to the simulation results is $\psi =
\psi_r \psi_p$.

\end{appendix}

\end{document}